\documentclass{emulateapj}
\usepackage{apjfonts}
\usepackage{graphicx}
\usepackage{psfig}

\newcommand{\begit}{\begin{itemize}}
\newcommand{\enit}{\end{itemize}}
\newcommand{\begen}{\begin{enumerate}}
\newcommand{\enen}{\end{enumerate}}

\newcommand       \be           {\begin{equation}}
\newcommand       \ee           {\end{equation}}
\newcommand       \bea          {\begin{eqnarray}}
\newcommand       \eea          {\end{eqnarray}}

\newcommand       \kms		{\,{\rm km \,\, s}^{-1}}
\newcommand       \cm		{\,{\rm cm }}
\newcommand       \pc		{\,{\rm pc }}
\newcommand       \yr		{\,{\rm yr }}
\newcommand       \s		{\,{\rm s }}
\newcommand       \dynes	{\,{\rm dynes }}
\newcommand       \g		{\,{\rm g }}
\newcommand       \kpc		{\,{\rm kpc }}
\newcommand       \K		{\,{\rm K }}
\newcommand       \M		{\,{\cal M }}
\newcommand       \erg		{\,{\rm erg }}
\newcommand       \ergs		{\,{\rm erg \,\, s}^{-1}}

\newcommand       \rcl		{r_{cl}}
\newcommand       \mcl		{M_{cl}}
\newcommand       \mcli		{M_{cl,i}}
\newcommand       \mj		{M_{J}}

\newcommand{\beqa}{\begin{eqnarray}} 
\newcommand{\eeqa}{\end{eqnarray}}

\begin{document}

\title{THE SIZES AND LUMINOSITIES OF MASSIVE STAR CLUSTERS}

\author{Norman Murray\altaffilmark{1,2}}

\altaffiltext{1}{Canadian Institute for Theoretical Astrophysics, 60
St.~George Street, University of Toronto, Toronto, ON M5S 3H8, Canada;
murray@cita.utoronto.ca} 
\altaffiltext{2}{Canada Research Chair in Astrophysics}

\begin{abstract}
The masses of star clusters range over seven decades, from ten up to
one hundred million solar masses. Remarkably, clusters with masses in
the range $10^4M_\odot$ to $10^6M_\odot$ show no systematic variation
of radius with mass. However, recent observations have shown that
clusters with $\mcl\gtrsim 3\times10^6M_\odot$ do show an increase in
size with increasing mass. We point out that clusters with
$\mcl\gtrsim 10^6M_\odot$ were optically thick to far infrared
radiation when they formed, and explore the hypothesis that the size
of clusters with $\mcl\gtrsim3\times10^6M_\odot$ is set by a balance
between accretion powered radiation pressure and gravity when the
clusters formed, yielding a mass-radius relation $\rcl\sim
0.3(\mcl/10^6M_\odot)^{3/5}\pc$. We show that the Jeans mass in
optically thick objects increases systematically with cluster mass. We
argue, by assuming that the break in the stellar initial mass function
is set by the Jeans mass, that optically thick clusters are born with
top heavy initial mass functions; it follows that they are
over-luminous compared to optically thin clusters when young, and have
a higher mass to light ratio $\Upsilon_V=\mcl/L_V$ when older than
$\sim 1$ Gyr. Old, optically thick clusters have $\Upsilon_V\sim
\mcl^{0.1-0.3}$. It follows that $L_V\sim\sigma^{\beta}$, where
$\sigma$ is the cluster velocity dispersion, and
$3.6\lesssim\beta\lesssim4.5$.  It appears that $\Upsilon_V$ is an
increasing function of cluster mass for compact clusters and
ultra-compact dwarf galaxies. We show that this is unlikely to be due
to the presence of non-baryonic dark matter, by comparing clusters to
Milky Way satellite galaxies, which are dark matter dominated. The
satellite galaxies appear to have a fixed mass inside a fiducial
radius, $M(r=r_0)=const.$.

\end{abstract}

\keywords{galaxies: star clusters --- stars: mass function}

\section{INTRODUCTION}
Most stars in the Milky Way are believed to have formed in star
clusters, e.g., \citet{L295}.  In their review \citet{Clarke} note that
$96\%$ of stars in the Orion B cloud are in clusters, while $50-80\%$
of stars in Orion A are. The latter fraction is consistent with that
found by \citet{1993AJ....105.1927G} in
Tarus. \citet{2007prpl.conf..361A} suggest that only $\sim20-25\%$ of stars
form outside clusters.  

There is independent evidence, based on simulations of field binary
star synthesis rather than on star counts in star forming regions,
that the majority of stars formed in clusters
\citep{1995MNRAS.277.1491K}. This result applies to stars formed over
the lifetime of the Milky Way disk. The similarity between the stellar
initial mass function (IMF) inferred from field stars and that
measured in young clusters implies that most low mass stars form in
clusters, and hence along side massive stars, rather than in an
isolated mode, at least in the Milky Way.

Observations of nearby galaxies suggest that a significant fraction of
stars form in clusters in these galaxies as well. For example,
\citet{meurer95} find that $20\%$ of the UV light in seven starburst
galaxies they surveyed comes from young star clusters. Similarly, the
cluster fraction of the B band light in the galaxy merger NGC 3256
found by \citet{zepf99} was $19\%$, while \citet{2005ApJ...631L.133F}
found that $20\%$ of the total H$\alpha$ emission in the Antennae
galaxies comes from clusters.

In the case of these external galaxies, the estimates of the cluster
fraction ($\sim20\%$) are likely to be lower limits, a point noted by
\citet{2005ApJ...631L.133F}. For example, \citet{meurer95} estimated
that the cluster fraction of young stars in NGC 5253 was $\sim14\%$
since that fraction of the 2200 \AA\ UV light came from
clusters. However, more recent radio and infrared observations of NGC
5253 have revealed the presence of a deeply embedded star cluster,
with a luminosity $\sim 4\times10^{42}\ergs$
\citep{2003Natur.423..621T}, and a correspondingly large ionizing flux
$Q\approx  7\times10^{52}\s^{-1}$ \citep{2004ApJ...602L..85T}
indicating an age of order 1 Myrs or less. This is $43\%$ of the
bolometric luminosity of the galaxy, given by
\citet{2001ApJ...554L..29G} 
as $9.2\times10^{42}\ergs$, showing that {\em at least} $43\%$ of the
star formation in this galaxy occurs in a single cluster. This
extremely luminous cluster was not detected in the UV by
\citet{meurer95}. It seems likely that much of the clustered star
formation in starburst galaxies is similarly obscured, so that the
clustered star formation estimate of \citet{meurer95} is significantly
low.

Observations of young star clusters find a power
law mass distribution of the form
\be  \label{eqn:mass distribution} 
{dN_{cl}(m)\over dm}=N_0m^{-\alpha}
\ee   
with $1.5<\alpha<2$ in both the Milky Way and in other galaxies
\citep{1989ApJ...337..761K,1997ApJ...476..144M}. This implies that
most massive stars are found in the few most massive clusters. If one
believes that massive stars are relevant to galaxy formation,
understanding star formation in massive clusters is crucial.

In this paper we show that clusters with initial
$\mcl\gtrsim3\times10^6M_\odot$ were likely supported by radiation
pressure just before and during the time their stars were forming; it
follows that such clusters will exhibit a mass radius relation
$\rcl\sim \mcl^{3/5}$, unlike globular clusters, which have radii
independent of their masses (by cluster radius we mean the projected
radius that encloses half of the cluster light).  

In addition, we argue that star formation in massive
($\mcl\gtrsim10^6M_\odot$) clusters will produce an IMF with a larger
characteristic mass than star formation in less massive clusters. The
IMF in the Milky Way is described by a broken power law
\citep{2001MNRAS.322..231K,2002ApJ...573..366M}, or by a log-normal
distribution \citep{1979ApJS...41..513M}. Similarly, the IMF in young
Milky Way clusters such as Orion is described by a broken power, e.g.,
\citep{2002ApJ...573..366M}, similar to that of the Milky Way as a
whole.  While the classic \citet{1955ApJ...121..161S} IMF has only an
upper and lower mass cutoff, these more recent power-law based
estimates of the IMF involve at least one additional characteristic
mass. In either the log-normal or powerlaw models, the characteristic
mass in the Milky Way is of order $0.5M_\odot$.

The origin of the characteristic mass is disputed in the
literature. For example, \citet{adams} state that ``\ldots the Jeans
mass has virtually nothing to do with the masses of forming
stars''. In contrast, \citet{McKee07} state ``A recurrent theme in
star-formation theory is that the characteristic mass--defined by the
peak of the IMF--is the Jeans mass at some preferred density.'' We
argue that the second point of view is correct.

The gas in the interstellar medium is not smoothly distributed; a
substantial fraction is found in the form of giant (tens of parsec
size and $10^6M_\odot$ mass, in the Milky Way) molecular clouds, which
contain parsec scale clumps, which in turn contain sub-parsec scale
cores; see, e.g., \citet{McKee07,bergin07}. The density increases as
one moves down the size and mass scale. The masses typically follow a
power law distribution similar in form to that of massive star
clusters (eqn. \ref{eqn:mass distribution} above). This behavior is
consistent with the notion that the density and mass distributions are
controlled by supersonic turbulence. Both analytic theory and
simulations predict log-normal density distributions in the gas density
and gas surface density of supersonic turbulent flows; observations
find surface densities that are consistent with log-normal
distributions.

Measured velocity differences are scale dependent, again consistent
with turbulent motions; they are large on large scales, but decrease
with decreasing scale. The length scale on which the velocity is equal
to the thermal sound speed is called the sonic length. Below this
scale turbulent motions cannot compress the gas, so that the density
is no longer controlled by turbulence; rather, thermal or magnetic
pressure, along with the self-gravity of the gas, will control the
density. In the Milky way the sonic length is $\sim0.03\pc$. Star
forming cores are typically about this size or smaller.

We assume that on scales below the sonic length the fragmentation
properties are controlled by the thermal properties of the gas.  For
an ideal gas equation of state $P=\rho kT/\mu\sim \rho^\gamma$, where
$\gamma$ is the effective adiabatic index of the gas. The Jeans mass
scales as $T^{3/2}/\rho^{1/2}\sim \rho^{(3\gamma-4)/2}$. If
$\gamma<4/3$ an increase in the gas density (as, for example, when the
clump is self-gravitating) leads to a decrease in the Jeans mass and
the likelihood of further fragmentation.  When $\gamma>4/3$ the Jeans
mass is an increasing function of density, and so in the absence of
other physics the gas will not fragment. For an isothermal equation of
state $\gamma=1$, while for an adiabatic equation of state
$\gamma=5/3$.

A hard lower bound to the characteristic mass is given by the opacity
limit \citep{LLB,Rees}, expressed by the condition that the accretion
luminosity of a contracting sphere of gas not exceed the black body
luminosity of a sphere of the same radius. The critical mass in this
case is just the Jeans mass, at the radius and density when the sphere
becomes optically thick. This bound applies whether the protostellar
core is formed by turbulence and transient, or if it
self-gravitating. In either case the luminosity is given roughly by
$L\sim v^5/G$; the velocity is $v\sim\sqrt{GM/r}$ if the clump is
contracting, and larger if it is transient.

Other theories associate the characteristic mass with the Jeans mass
at some other density, e.g., with the density at which the gas first
becomes thermally coupled to dust grains
\citep{larson73,larson05,jappsen05}. The argument is that the gas is
nearly but not quite isothermal; at low density the temperature
decreases with increasing density, but above the critical density
temperature begins to increase with increasing density. This is
modeled as a change in the effective adiabatic index $\gamma$ of the
gas, from less than unity to slightly more than unity; numerical
experiments \citep{li03} indicate that gas fragments readily when
$\gamma<1$, but less readily for $\gamma>1$. 

\citet{2006MNRAS.368.1296B} perform numerical experiments, using an
isothermal equation of state, that show that the mass of the break in
the IMF is equal to the initial Jeans mass, and increases with
increasing initial Jeans mass. When they employ a
Larson style equation of state, they find a fixed characteristic mass
in the resulting IMF, corresponding to the density where the adiabatic
index suddenly increases from $\gamma<1$ to $\gamma>1$.

Competitive accretion \citep{zinnecker82,bonnell01} schemes also
predict that the characteristic mass is the mean Jeans mass, i.e., the
Jeans mass using the mean density of the star forming region, and
$T\approx10$ K \citep{1998ApJ...501L.205K,2006MNRAS.368.1296B}.

Yet other theories for the origin of the IMF start from the notion
described above, that the observed fragmentation in giant molecular
clouds arises from turbulence, e.g., \cite{PN}. The turbulence
establishes a powerlaw in clump mass at the high mass end, with an
index similar to the Salpeter values. However, even in this work the
{\em characteristic mass} is related to the Jeans mass (calculated
using the mean density) divided by the Alfvenic or fast magnetosonic
Mach number ${\cal M}_F$ to the $2/3$ power \citet{PN}, their equation
30.

\citet{elmegreen08} note that numerical simulations consistently show
a proportionality between the characteristic mass and the thermal
Jeans mass, and point out that the observed constancy of the
characteristic mass then requires a constant Jeans mass, despite
variations in the environments where stars from. We argue below that
under conditions found in ultraluminous infrared galaxies (ULIRGs) and
other massive galaxies neither the Jeans mass nor the characteristic
mass is like that found in the Milky Way.

This paper is organized as follows. In \S\ref{sec: Cluster} we show
how the Jeans mass $\mj$ varies with cluster mass. When the star
forming clumps have $\rcl\sim1\pc$ and $\mcl\gtrsim10^6M_\odot$, the
finite optical depth to the far infrared radiation released by the
contraction of the protocluster gas or by the dissipation of turbulent
motion leads to an increase in temperature, and, we show, in the Jeans
mass. The role of radiation pressure in setting the radius of very
massive clusters is described in \S\ref{sec:radsupport}. In
\S\ref{sec:compare} we compare our results to observations of massive
young clusters, of ultra-compact dwarfs (UCDs) and of central massive
objects such as those in \citet{2002AJ....124.3073G} and
\citet{walcher05}. We give a short discussion of our results in the
context of previous models for very massive star clusters in
\S\ref{sec:discussion}, and offer our conclusions in
\S\ref{sec:conclusions}.  In the appendices 
we discuss the log-normal distribution, the initial mass
function (or IMF) that we use and the calculation of the associated
light to mass ratio, we outline the calculation of the mass-radius
relation for a radiation pressure supported cluster, and finally we
discuss the dark matter density on $10\pc$ scales predicted by recent
numerical simulations.

\section{PROTOCLUSTER FRAGMENTATION AND THE INITIAL MASS FUNCTION}
  \label{sec: Cluster} 

Observations of embedded clusters in the Milky Way show that the star
formation efficiency (SFE), the fraction of cluster gas that ends up
in stars, is $10-30\%$ \citep{L2} for clusters with stellar mass $\sim
10-1000M_\odot$. In contrast to this well established result, the time
scale over which the gas is converted into stars is nearly as
contentious as the the origin of the characteristic mass; it is either
the dynamical time, e.g., \citet{elmegreen2000}, or a few to several
dynamical times, e.g., \citet{tkm}. In appendix
\ref{appendix:lognormal} we show that the combination of relatively
high SFE and short star formation timescale implies that the initial
density of the gas that ends up in stars is no more than a factor
$\rho_{m}/\bar\rho\sim10$ larger than the mean density $\bar\rho$ of
the cluster for rapid star formation; it follows that the Jeans mass
of this gas is no less than $M_J/<M_J>\sim 1/3$, where $\langle
M_J\rangle$ is the Jeans mass calculated using the mean cluster
density. For a more extended star-formation time, say $5$ dynamical
times, and the minimum SFE of $10\%$, this ratio may be as low as
$1/6$. Both these estimates are likely to be low compared to reality,
since some stars undoubtedly form in less dense but more extended
self-gravitating sub-clumps.

The point here is that, while log-normal distributions are broader than
gaussians, in the present context they are still rather narrow, so the
gas destined for star formation cannot come from too far out on the
tail of the density distribution.

Observations of magnetic fields, via Zeeman splitting in OH masers,
in both the Milky Way, with $B\sim 3{\rm \ mG}$ \citep{fish}, and in
nearby ultraluminous infrared galaxies, also with $B\sim 3{\rm\ mG}$
\citep{robishaw} indicate that the fast magnetosonic Mach number
${\cal M}_F\lesssim 5$. \citet{thompson06} argue that in ULIRGs the
volume average field is also $\sim1{\rm\ mG}$, so that the Mach number
is globally of the same order.

Simulations suggest that $\rho_m/\bar\rho$ depends only weakly on
$\M_F$ of the turbulence \citep{osg01}. Since the Mach number
does not vary over a large range, and assuming that turbulence in the
ISM of galaxies is adequately modeled by recent simulations, it
follows that the mean Jeans mass of the cluster is a good proxy for
the mean Jeans mass of the star forming gas in clusters.

Henceforth we will assume that the characteristic mass is in fact
the Jeans mass of the proto-stellar gas,
\be  \label{eqn:Jeans mass}
M_J=\phi_J\left({\lambda_J\over 2}\right)^3\bar\rho,
\ee  
where
\be  
\lambda_J\equiv\sqrt{\pi c_s^2\over G\bar\rho}
\ee  
is the Jeans length,  $c_s^2=kT/\mu m_p$ is the sound speed, and
$\phi_J$ is a dimensionless constant of order unity, accounting for
the difference between the mean density of the cluster and the mean
density of that fraction of the cluster gas that ends up forming stars.

In a proto-cluster clump of gas, the mean density is given by
\be  
\bar\rho={3\mcl\over 4\pi \rcl^3},
\ee  
where $\mcl$ and $\rcl$ are the initial gas mass and radius. 

We note that it is likely that both the initial $\mcl$ and $\rcl$
differ from the final cluster mass and radius. 
For example, we noted that the SFE was of order $10-30\%$.
If the unused gas is expelled on
a dynamical time or longer, the cluster will expand by a factor $\sim
1/SFE$ \citep{Hills} from its original radius. This result is
confirmed by numerical simulations (\citet{Baumgardt_Kroupa}, see
their Fig.~4). The simulations show that if the gas is expelled on a
shorter time scale, the cluster will expand by a somewhat larger
factor, or it may even be destroyed, if the SFE
is less than $30\%$. \cite{Baumgardt_Kroupa} show that the final
cluster radius will actually be smaller than the initial radius if the
cluster is subject to strong tidal forces from its host galaxy.

It is not clear that a cluster with $\mcl=10^6M_\odot$ and $\rcl=3\pc$
will have SFE as low as $\approx
0.3$. Such a cluster has an escape velocity of order $50\kms$, well
above the sound speed of photoionized gas, so any HII region that is
formed will be dynamically irrelevant. To see this, note that the
dynamical pressure in the cluster is 
\be  
P\approx\pi G \Sigma^2\approx8\times10^{-6}\g\cm^{-1}\s^{-2}.
\ee  
Meanwhile, the mean particle number density is
$n\approx3\times10^5\cm^{-3}$, so the pressure of ionized gas is
$P_{HII}=4\times10^{-7}\g\cm^{-1}\s^{-2}$, dynamically negligible. 

Nor can stellar winds expel the cluster gas; the bolometric luminosity
of $10^6M_\odot$ of zero age main sequence stars is roughly
$L=6\times10^{42}\ergs$, while the wind kinetic luminosity is
$L_w\approx2\times10^{40}\ergs$. The cooling rate of the wind is
$L_{cool}=\Lambda n_h^2 V\approx 10^{42}\ergs$, where $V$ is the
volume of the cluster and $n_h\approx6\times10^3\cm^{-3}$ is found by
assuming the shocked stellar wind has a pressure comparable to the
dynamical pressure. In other words, the wind luminosity is not
sufficient to produce a pressure high enough to support the weight of
the overlying gas. Alternately, the cooling time
$\tau_{cool}=kT/\Lambda n_h\approx2\times10^{10}\s$ is much shorter
than the dynamical time $\tau_{dyn}=r/v\approx3\times10^{12}\s$; the
wind cools before it can push the overlying cold gas out of the
cluster.

Supernovae will not remove much gas from the cluster either; the
binding energy of the cluster is $3\times10^{52}\erg$, about $30$
times the kinetic energy supplied by a single supernova. While one
expects multiple supernovae to explode, they do not explode all at
once. For typical initial mass functions the supernova luminosity is
$L_{SN}\approx10^{40}\ergs$, similar to the wind luminosity. Thus
supernovae deposit a total energy $L_{SN}\tau_{dyn}\approx
2\times10^{50}\erg$ in a dynamical time, not sufficient to unbind the
cluster (of course the gas cools in less than a dynamical time).

\begin{figure}
\resizebox{\hsize}{!}{\includegraphics{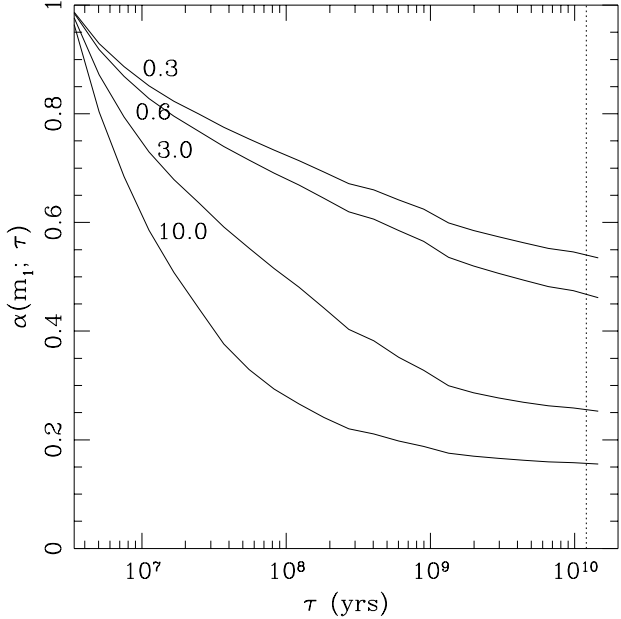}}
\caption{
  The lockup fraction $\alpha(m_1,z/z_\odot=0.2;t)$ for the IMF given in
  equation (\ref{eq: muench}) and several values of $m_1$, as
  labeled. The vertical dotted line is at $12$ Gyrs.
\label{Fig:lockup}}
\end{figure}

This argument applies to young clusters, with a substantial fraction
of their mass in the form of gas. However, for clusters in which the
gas fraction is very small, gas removal may be very efficient. For
example, in our $10^6M_\odot$ cluster, if the gas fraction is reduced
to $1/30$ of the total mass (perhaps because the rest of the mass has
been put into stars), the binding energy of the gas is reduced to
$\sim10^{51}\erg$.  In that case, a single supernova may be able to
eject the gas. Some such mechanism appears to operate in Milky Way
globular clusters, as inferred via the following argument: as stars
evolve off the main sequence, they lose their envelopes.  The mass
loss rate from this process would lead to substantial amounts of
intercluster gas accumulating, of order $100-1000M_\odot$ between
passages through the plane of the galaxy
\citep{1988IAUS..126..411R}. Observations of the intercluster medium
in globular clusters \citep{2001ApJ...557L.105F} show conclusively
that this gas does not remain in the cluster, resulting in a net
decrease in cluster mass.  This reduction in cluster mass is
quantified in the so-called lockup fraction $\alpha(t)$, the fraction
of mass locked up in present day stars and stellar remnants; see,
e.g., chapter 7.3 of \citet{Pagel}. For a Muench et al. IMF with
$m_1=0.6M_\odot$, $\alpha(12Gyrs)\approx 0.45$, see Figure
\ref{Fig:lockup}. In our context, $\alpha(t)$ is a very rough lower
limit to $\epsilon$.

The Jeans mass associated with a cluster of mass $\mcl$ is set by the
temperature $T$ and the radius $\rcl$. We examine the Jeans mass in
three different cases; 1) when the cluster is either optically thin to
far infrared radiation (FIR), or of sufficiently low accretion
luminosity that the effective temperature is $\sim10\K$; 2) when the
cluster is optically thick and $T_{eff}>>10\K$, but the cluster is not
radiation pressure supported, and 3) when the protocluster gas is
radiation pressure supported by the accretion generated luminosity.

To be quantitative, we need to pick a particular IMF. We will use a
modified version of the \citet{2002ApJ...573..366M} IMF, described in appendix
\ref{appendix:imf} and quantified in equation (\ref{eq: muench}). We consider two
modifications to the Muench et al. IMF. The first consists of ignoring
the small mass bump (between masses of $0.025$ and $0.017M_\odot$),
extending the $0.73$ power law down to the minimum mass we consider,
usually $m_L=0.1M_\odot$. The second modification is that we use a
maximum stellar mass of $M_U=120M_\odot$. Since we ignore the
small-mass bump, there are two free parameters (in addition to $m_L$
and $m_U$) in our IMF, the two break masses $m_1$ and
$m_2<m_1$. Muench et al. find that in the Orion Nebular Cluster,
$m_1=0.6M_\odot$ and $m_2=0.12M_\odot$. In this paper we assume that
the larger mass is in fact the Jeans mass, $m_1=M_J$.

\subsection{Optically thin protoclusters}
When the protocluster is optically thin to far infrared radiation, as
is the case for almost all clusters in the Milky way at the present
epoch, the temperature is very nearly independent of cluster radius
(or alternately, density); the gas temperature is observed to be
around $10-20K$ e.g., \citet{1988ApJS...68..257C,rathborne06}. The Jeans
mass is then only dependent on the cluster density. If the mass-radius
relation in young Milky Way clusters is taken to be
$\rcl\sim\mcl^{1/3}$, which is consistent with the clusters listed in
\citet{L2}, the Jeans mass is the same in all current Milky
Way protoclusters, and hence (under our assumptions) so is the
IMF. This is consistent with observations of the IMF in the Milky Way,
except possibly in the galactic center.

For example, consider the best studied young cluster, the
Trapezium-ONC cluster in Orion. \citet{Hillenbrand} lists 1576 stars,
of which 973 are given a probability larger than $0.5$ of being
members; Hillenbrand gives a lower limit to the stellar mass of
$900M_\odot$, which she estimates is about half of the actual
mass. Half the stars lie within $0.72\pc$ of the Trapezium, assuming a
distance of $400\pc$ to Orion \citep{2005A&A...438.1163K}. The
extinction in our direction is small, $A_v<2.5$, while the surface
density of stars is currently
$\Sigma=1800M_\odot/1.7\pc^2=0.2\g\cm^{-2}$, corresponding to
$N_H\approx 10^{23}\cm^{-3}$ and $A_v\approx 70$.

As noted above, the star formation efficiency of embedded clusters in
this mass range is of order $30\%$ \citep{L2}. This suggests that the
protocluster which formed the ONC had a mass $\sim 3$ times
larger than the current stellar mass. The original dynamical time of
the protocluster was $R/v\sim 10^5(R/0.7\pc)(4\kms/v)\yr$, or less, as
the original radius was likely smaller than the current radius.

If the gas was removed when it was ionized by the central O stars, its
outflow velocity might have been of order $10\kms$, (the sound speed
$c_s\approx 13\kms$) but the pressure of the ionized gas has to
overcome the weight of the bulk of the cluster gas. The pressure of
the gas was $\pi G\Sigma^2\approx 10^{-8}\dynes\cm^{-2}$, while the
pressure of the ionized gas is
\be
\begin{array}{l}  
P_{HII}=\sqrt{3Q\over 4\pi\alpha_{rec} r^3}kT\approx 10^{-9}
\left({Q\over 2\times10^{49}\s^{-1}}\right)^{1/2}\\
\quad\left({0.72\pc\over r}\right)^{3/2}
\dynes\,\cm^{-2},
\end{array}  
\ee 
where $Q$ is the number of ionizing photons emerging from the O stars
per second, and $\alpha_{rec}$ is the recombination coefficient; we assume
photoionization equilibrium. It would appear that initially the
ionized gas cannot disrupt the protocluster gas; protostellar jets are
a promising candidate for dissipation of cluster gas, but the rate of
momentum deposition is such that the removal time is rather long. If
so, it is likely that the gas removal was, very roughly speaking,
adiabatic.

The original radius of the cluster would then have
been $r\approx 0.24\pc$, while the original mass was $M\approx 3\times2
\times900M_\odot=5400M_\odot$. The mean density was
$\rho\approx 6\times10^{-18}\g\cm^{-3}$. The Jeans length
$\lambda_J=8\times10^{16}\cm$, and the Jeans mass $M_J\approx 
4\times10^{32}\g$, or $0.2$ solar mass, similar to the value of
$m_1=0.6M_\odot$ found by \citet{2002ApJ...573..366M}.

The surface density of the proto-ONC gas was 
\be  
\Sigma\equiv {M\over 4\pi r^2}\approx  1.6
\left({M\over 5400M_\odot}\right)
\left({0.24\pc\over r}\right)^2
\g\cm^{-2}
\ee  
as seen from the center of the cluster. This is sufficiently high that
we must consider the possibility that the clump was optically thick to
far infrared radiation. If we assume $T=20K$ is correct, using the
Rosseland mean opacity
\be  
\kappa(T)=3\left({T\over 100K}\right)^2\cm^2\g^{-1}
\ee  
\citep{2003A&A...410..611S} we find $\tau=0.2$. Hence the proto-ONC
cloud was optically thin, but only just. 

If, on the other hand, we assume the cluster was
optically thick, and calculate the accretion luminosity (as is done in
the next subsection) we obtain $T_{eff}=14$K, and we still find an optical
depth less than unity; we conclude that the ONC was optically thin.

\subsection{Optically thick protoclusters lacking radiation pressure
  support} \label{sec: thick} 
The situation will change if the protocluster is optically thick to
far infrared radiation, and if the accretion luminosity is
sufficiently high. The temperature of the gas will then be higher
than in optically thin clusters, potentially leading to a larger Jeans
mass in optically thick clusters. For a metallicity $Z$ gas the
optical depth is
\be  
\begin{array}{l}
\tau=\kappa(T,Z)\Sigma_{cl}
\approx 1
\left({M_{cl}\over 2\times  10^5M_\odot}\right)
\left({1\pc\over \rcl}\right)^{-2}\\
\quad
\left({\kappa(100,Z_\odot)\over 3\cm^2\g^{-1}}\right)
\left({T\over 100}\right)^2
\left({Z\over 0.1 Z_\odot}\right),
\end{array}
\ee  
where $\kappa(T,Z)$ is the Rosseland mean opacity, and we have used
the fact that $\kappa(T)\sim(T/100 K)^2$ for $T\lesssim100K$; for
$100K<T<500K$, $\kappa$ is roughly constant.

We have scaled to the properties of a ``metal rich'' (one tenth solar)
Milky Way globular cluster, using a typical present day (stellar) mass
but a small half light radius. The reason for the latter choice is
that the lockup fraction $\alpha(t)\approx 0.45$; if this mass was
lost from the cluster, the radius of the cluster has expanded by a
factor of $2.2$ since it formed.  We noted above that observations of
the intercluster medium in globular clusters, e.g.,
\citet{2001ApJ...557L.105F}, show that the gas expelled from evolving
stars is rapidly removed from the cluster.

If the star formation was less than $100\%$ efficient, the amount of
expansion would have been larger.

The heat source that, for massive or compact enough clusters, drives
the gas above $T\approx 10K$ is the contraction of the clump due to its own
self-gravity. Consider a clump of mass $\mcl$ and
radius $\rcl$, contracting at roughly its free fall
time $\tau_{ff}$. The cluster generates a luminosity
\be 
L_{acc}={G\mcl^2\over \rcl \tau_{ff} }=\phi_{ff}{v^5\over G},
\ee 
where $\phi_{ff}$ is a dimensionless constant of order unity, and 
\be %
v=\sqrt{G\mcl\over\rcl}.
\ee %

The luminosity of a $10^6M_\odot$ cluster with
$\rcl=1\pc$ contracting at the free fall rate is 
%
\be  
L_{acc} \approx 
1.9\times10^{41}
\phi_{ff}
\left({M_{cl}\over 10^6M_\odot}\right)^{5/2}
\left({\rcl\over 1\pc}\right)^{-5/2}
\ergs.
\ee %
We stress once again that this luminosity is not related to any star
formation.

The effective temperature of the collapsing gas is
\be \label{eqn:T_eff}%
T_{eff}\approx  17\ \phi_{ff}^{1/4}
\left({\mcl\over 10^5 M_\odot}\right)^{5/8}
\left({\rcl\over 1\pc}\right)^{-9/8}
\K.
\ee %
For $\mcl=10^5M_\odot$ this is comparable to the temperature seen in
star forming cores in the Milky Way, so we would not expect any change
in the IMF. 

However, for clusters larger than $\sim10^6M_\odot$, $T\approx70K$,
and the interior temperature of the protocluster is higher still.
To find the run of $T$ with $r$, we solve the radiative transfer
equation in the diffusive limit, employing the Rosseland mean opacity
as calculated by \citet{2003A&A...410..611S}. We start at the surface
of the clump, with a known luminosity, radius, and effective
temperature. Since the clump is nearly in free-fall, we assume the
density scales as $\rho(r)\sim 1/r^2$. With $\rho$ and $T$ known, we
can find the opacity. Integrating inwards, we solve for the
temperature; then using our prescribed density and the new $T$, we
find the opacity, and proceed inwards.

Once we know $T(r)$, $\kappa(r)$, and (our assumed) $\rho(r)$, we
calculate a mass weighted temperature. For the parameters used here,
we find $T_{mass}\approx 165\K$, but this varies with the mass and
metallicity of the cluster

The corresponding Jeans length is
\be \label{eq:jeans length}%
\lambda_{Jeans} 
\approx 1.4\times10^{17}
\cm,
\ee %
while the Jeans mass is
\be \label{eqn: Jeans mass thick}%
M_{Jeans} 
\approx 2.7
\phi_{J}
M_\odot.
\ee %

\begin{figure}
\plotone{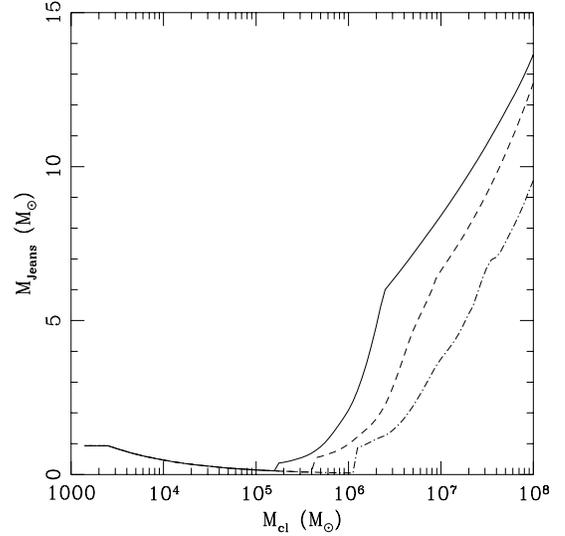}
\caption{
  The Jeans mass plotted against stellar cluster mass $\mcl$;
  $\mcl$ is the initial gas mass of the proto-cluster, i.e., the plot
  does not take account of any gas expelled from the cluster as the
  stars are formed. The metallicities are $z/z_\odot=1$ (solid line),
  $z/z_\odot=0.1$ (dashed), and $z/z_\odot=0.01$ (dot-dash). The Jeans
  mass initially decreases with increasing $\mcl$, since $T$ is
  constant while $\rho$ increases (because we have assumed $\rcl$ is
  constant for $2000<\rcl<10^5M_\odot$); this decrease is halted when
  the cluster becomes optically thick, leading to an increase in $T$
  and hence in $\mj$. The result is a rapid increase in $\mj$ with
  increasing $\mcl$. There is another change in behavior when the
  protocluster becomes radiatively supported, for
  $\mcl\gtrsim10^6M_\odot$ (depending on metallicity). See text for
  further details.
\label{Fig:Jeans_Mass_cl}}
\end{figure}

Figure \ref{Fig:Jeans_Mass_cl} plots $\mj$ as a function of $\mcl$ for
three different values of the metallicity. The cluster radius is set
to $1.0\pc$ for $2000M_\odot<\mcl<\mcl^*$, where
$\mcl^*\approx 10^6M_\odot$ is the mass at which $\rcl$ begins to rise
as a result of the increased accretion luminosity, as discussed below
in \S\ref{sec:radsupport}. For smaller masses
$\rcl$ is chosen to match the present day mean cluster radius
in the Milky Way, allowing for cluster expansion as the lockup
fraction $\alpha(t)$ decreases due to stellar evolution.  

The Figure shows that $\mj$ first decreases with increasing $\mcl$. At
small cluster masses, where the cluster is optically thin to FIR
radiation, the temperature is independent of $M_{cl}$, while the
density increases, so $M_J$ decreases with increasing $M_{cl}$. This
is a result of our assumption that the initial cluster radius does not
vary with cluster mass. This assumption can be checked by measuring
the IMF in globular clusters, and seeing if massive clusters show
signs of a sharp decrease in the number of stars, binned by mass, as
the stellar mass increases. This signature of the IMF may, however, be
erased by the combination of mass segregation and preferential
evaporation of low mass stars, e.g., \citep{2003MNRAS.340..227B}.

The Jeans mass rather abruptly jumps up (at $\mcl\sim
10^5-10^6M_\odot$, depending on metallicity); this jump is followed by
a rapid but smooth increase. The jump reflects the fact that the
cluster has become optically thick; it is artificially abrupt due to
our crude treatment of the radiative transfer when the optical depth
is near unity. The smooth increase reflects the increase in $T$ driven
by the increase in accretion luminosity with increasing $\mcl$, $T_{eff}\sim
\mcl^{5/8}$. Assuming a constant $\rcl$, we find $\mj\sim \mcl^{7/16}$.

There is a slight break in the slope at $\mcl\approx 10^6M_\odot$
(for solar metallicity) related to the increase in $\rcl$ mentioned
above and driven by radiation pressure support; at higher $\mcl$ the
temperature actually decreases, although very slowly, $T_{eff}\sim
\mcl^{-1/20}$. However, $\mj$ continues to increase rather rapidly,
because the mean density actually decreases, $\rho\sim
\mcl^{-4/5}$. We find $\mj\sim\mcl^{13/20}$. 

We note that the Jeans mass increases rather strongly with
increasing cluster mass; this will result in a mass to light ratio
$\Upsilon_V$ that, for old clusters, will also increase with
increasing $\mcl$. We will return to this point below.

\subsection{Mass to light ratios of clusters with large Jeans masses}\label{sec:LM_Jeans}
\begin{figure}
\plotone{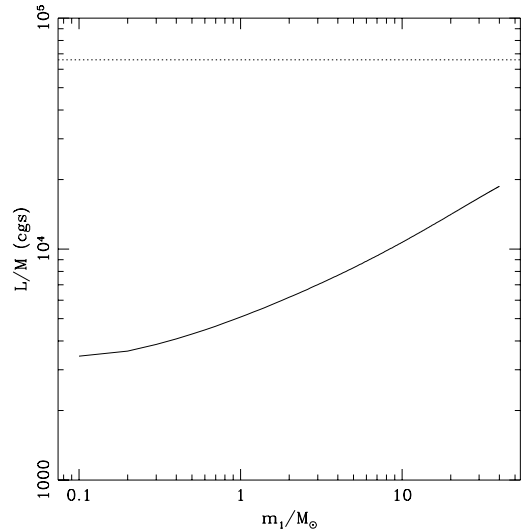}
\caption{
  The luminosity to mass ratio for a modified Muench et
  al. IMF, plotted as a function of $m_1$, interpreted as a Jeans
  mass, i.e. $m_1=m_J$. The cluster age is $2.5$ Myrs. The horizontal
  dotted line gives the Eddington ratio $4\pi
  Gc/\kappa_{es}=6.6\times10^4 \cm^2\s^{-3}$.
\label{Fig:Jeans_IMF}}
\end{figure}

The luminosity to mass ratio of the modified IMF will differ
significantly from that of the standard Muench et al. IMF; a larger
fraction of the resulting young stellar cluster is in massive stars,
leading to a higher luminosity to mass ratio. The luminosity to mass
ratio as a function of the Jeans mass is shown in
Fig. \ref{Fig:Jeans_IMF} for a cluster with an age of $2.5$ Myrs. The
horizontal dashed line is the Eddington luminosity to mass ratio,
$4\pi Gc/\kappa_{es}$, where $\kappa_{es}\approx  0.38\cm^2\g^{-1}$ is
the electron scattering opacity.

\begin{figure}
\plotone{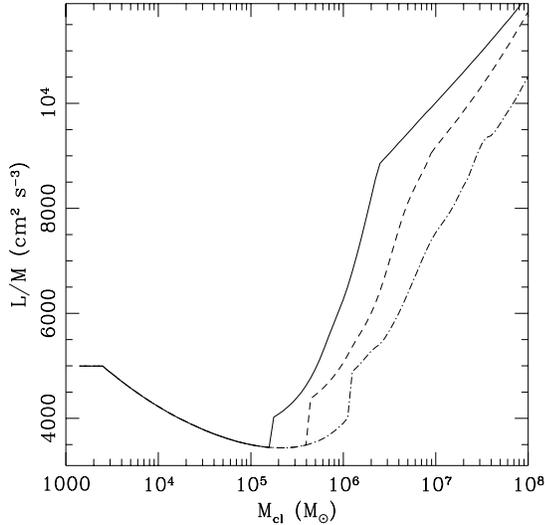}
\caption{Fig. 4---The luminosity to mass ratio for the clusters shown in Figure
  \ref{Fig:Jeans_Mass_cl}, at an age of $2.5$ Myr. The jump in $L/M$
  occurs at the mass where the cluster becomes optically thick to the
  far infrared radiation.
\label{Fig:L_over_M_Jeans_cl}}
\end{figure}

Figure \ref{Fig:L_over_M_Jeans_cl} shows the luminosity to mass ratio
for the same cluster models shown in
Fig.~\ref{Fig:Jeans_Mass_cl}. Recall that these models had $\rcl=1\pc$
for $10^4M_\odot<\mcl<10^6M_\odot$. It is not clear that this is the
correct mass-radius relation to use. However, the general trend of a
slowly varying $L/M$ ratio for small mass clusters, with an increase
for masses above some value ($\sim10^5M_\odot$ in the Figure) should
be correct for the actual mass-radius relation for massive
protoclusters.

\begin{figure}
\plotone{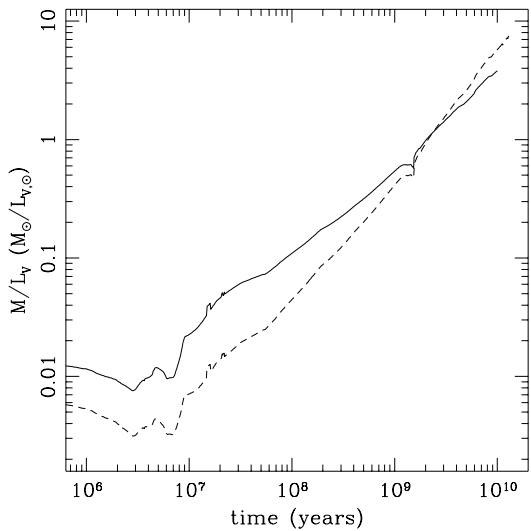}
\caption{
  The mass to light ratio $\Upsilon_V\equiv M/L_V$,
  in solar units (solar mass over solar v-band luminosity) for an
  optically thin (solid line, $m_J=0.5M_\odot$) and an optically
  thick (dashed line, $m_J=10.0M_\odot$) cluster as a function of
  cluster age, as calculated by Starburst99
  \citep{1999ApJS..123....3L,2005ApJ...621..695V}. The metallicities
  ($Z=0.2Z_\odot$) and initial stellar masses of the two clusters are
  the same. When the clusters are young, the optically thick cluster
  is more luminous than the optically thick cluster, but after a few
  gigayears the optically thin cluster is more luminous than the
  optically thick cluster.
\label{Fig:M_over_L vs age}}
\end{figure}

\begin{figure}
\epsscale{.9}
\plotone{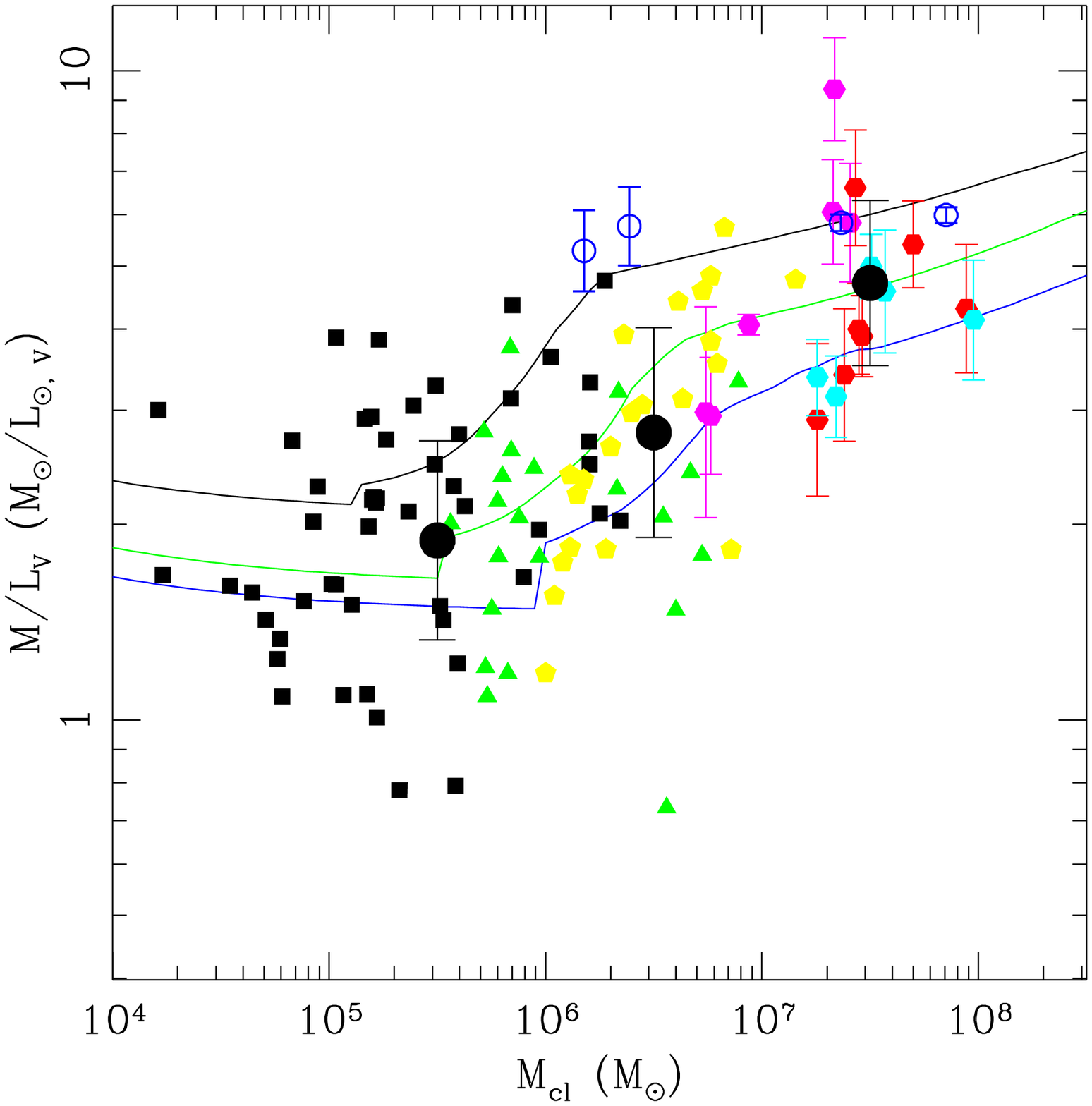}
\caption{
  The mass to light ratio $\Upsilon_V=M/L_V$ as a function of
  dynamical mass $\mcl$. The squares are Milky Way globular clusters
  (\citet{Harris}; velocities from \citet{pryor}). Triangles are M31 globular clusters
  (for which velocity dispersions are available) from
  \citet{barmby07}. The filled pentagons are globular clusters near
  NGC 5182 (Cen A) from \cite{Rejkuba}. The filled hexagons are UCDs
  from \citet{hasegan05} (magenta), Virgo UCDs from
  \citet{evstigneeva07} (red), and Fornax UCDs \citep{hilker07}
  (cyan). The open blue circles are the nuclei of Virgo dwarf elliptical
  galaxies from \citet{2002AJ....124.3073G}. The large filled circles
  are the mean $\Upsilon_V$ in three mass bins $\mcl<10^6 M_\odot$,
  $10^6M_\odot<\mcl<10^7M_\odot$, and $10^7M_\odot<\mcl$. The mass to
  light ratios of all clusters have been adjusted for the effects of
  two body relaxation and tides, as described in the text.
  The solid lines give the predictions of
  equations (\ref{eqn:Jeans mass}), (\ref{eqn:T_eff}), and (\ref{eqn:
  radiation length mass}) for a solar metallicity cluster (top curve),
  for a metallicity of $0.2$ solar (middle curve), and a metallicity
  of $0.02$ solar, all at an age of $10$ Gyrs.
\label{Fig:M_over_L vs M_cl}}
\end{figure}

While these optically thick clusters have high $L/M$ ratios compared
to optically thin clusters when they are young, they have low $L/M$
ratios, or high $M/L$ ratios, when they are more than a few Gyrs old,
as illustrated by the lines in Figs. \ref{Fig:M_over_L vs age} and
\ref{Fig:M_over_L vs M_cl}. The reason is clear; when young, the cluster
luminosity is dominated by massive stars, and optically thick clusters
have more massive stars per unit mass (and hence fewer low mass stars
per unit mass) than optically thin clusters. However, after all the
massive stars have evolved, the cluster light is supplied by low mass
stars, and the optically thin clusters have more $0.6-0.8M_\odot$ stars per
total mass than do the optically thick clusters; the optically thick
clusters have more stellar remnants (primarily white dwarfs, with some
neutron stars and possibly black holes) than the optically thin
clusters.

\section{RADIATION PRESSURE SUPPORTED PROTOCLUSTERS}
\label{sec:radsupport}
Another dramatic change occurs when the protocluster is massive enough
that the accretion luminosity is dynamically important. The
protocluster has an FIR optical depth larger than one, so the outward
force exerted by the radiation is
\be \label{eqn:radiation force}
F={\tau L_{acc}\over c}.
\ee 
If this force is less than that of gravity, the protocluster will continue to
shrink, but as $r$ decreases the radiation force increases as
$r^{-9/2}$, while the force of gravity increases as $r^{-2}$. For
small enough $r$ the radiation pressure overcomes the force of gravity
and the collapse is slowed.

The optical depth $\tau$ is measured from the center of the
protocluster outward. Equation (\ref{eqn:radiation force}) assumes
that the optical depth is independent of direction, which is not
strictly speaking true given the turbulent nature of the cluster
gas. The optical depth is proportional to the column density of
gas. The column density distribution has been measured in the Milky
Way by a number of authors, e.g., \citet{goodman} or \citet{wong}, and
has been found to be consistent with a log-normal
distribution. Numerical simulations also find log-normal surface
density distributions \citep{osg01}. The observations measure $\tau$
along the line of sight from the Earth through the cloud rather than
from the center of the cloud outward, but the two surface density
distributions should not differ dramatically. In the notation of
appendix \ref{appendix:lognormal}, \citet{goodman} find
$0.11<\sigma<0.22$, corresponding to $0.01<\mu<0.05$. This agrees well
with the results of \citet{osg01}.

For $\mu=0.05$, $99\%$ of sight lines have $\tau/\bar\tau>0.2$, where
$\bar\tau$ is the (angular) mean of the optical depth. Since
$\bar\tau\approx50$ for $\mcl=10^6M_\odot$ and $\rcl=1\pc$, there are
essentially no optically thin sight lines for such massive
clusters. In fact, the radiation pressure does not become dynamically
important until the cluster is smaller than $1\pc$ and $\bar\tau$ is
larger, but this only decreases the chance that photons leak out in
directions with small optical depth.

When the radiation force approaches that of gravity, the collapse will
slow from the free fall rate to the Kelvin rate. In appendix
\ref{appendix: size} we show that this occurs at a radius
\be \label{eqn: radiation length mass}
\begin{array}
r_{rad}=\phi_{rad}
G^{1/5}
\left({\kappa\over 4\pi c}\right)^{2/5}
\mcl^{3/5}
\approx 1.5\times10^{18}
\left({\mcl\over 10^6M_\odot}\right)^{3/5}\\
\quad
\left({\kappa\over 3\cm^2\g^{-1}}\right)^{2/5}
\left({\phi_{rad}\over 3}\right)
\cm.
\end{array}
\ee 
It can be shown that this is also the radius at which the photon
diffusion time out of the clump is equal to the clump dynamical time.

These radiation supported clusters have a monotonically increasing
$\rcl(\mcl)$ relation, unlike the less massive globular clusters found
in the Milky Way and other nearby galaxies. This will change a number
of their properties. For example, their surface densities will {\em
  decrease} with increasing mass,
\be \label{eqn:Sigma} 
\Sigma(\mcl)= 70
\left({\mcl\over 10^6 M_\odot}\right)^{-1/5}
\left({\kappa \over 3\cm^2\g^{-1}}\right)^{-4/5}
\left({\phi_{rad}\over 3 } \right )^{-2}\g\cm^{-2},
\ee  
where we use $\Sigma\equiv M/(4\pi r^2)$. This
scaling with cluster mass is in contrast to globular clusters, which
have surface densities that increase with $\mcl$, $\Sigma(\mcl)\sim
\mcl$. The volume density will decrease even more rapidly
\be  
\rho(\mcl)\sim \mcl^{-4/5}.
\ee  
This will have consequences for the evolution of binary systems, in
particular the precursors to LMXBs and millisecond pulsars. Since the
formation of such binaries depends strongly on the stellar number
density, the number of LMXBs and millisecond pulsars per stellar mass
will peak at a cluster mass $\sim3\times10^6M_\odot$.

The escape velocity will also
behave differently for the more massive clusters,
\be \label{eqn: escape} 
v(\mcl)\approx  93
\left( {\mcl\over 10^6 M_\odot} \right)^{1/5}
\left( {\kappa\over 3\cm^2\g^{-1}} \right)^{-1/5}
\left( {\phi_{rad}\over 3} \right)^{-1/2}
\kms
\ee  
compared to the much more rapid $v(\mcl)\sim \mcl^{1/2}$ for globular
clusters. 

It follows that the accretion luminosity $L\sim \mcl$ as expected
(assuming $\kappa$ is constant) since the cluster is limited by the
Eddington luminosity. Furthermore, the effective temperature will
actually decrease with increasing cluster mass, although very slowly,
$T_{eff}\sim\mcl^{-1/20}$.

The stellar luminosity is another question. When the stars ignite,
they will increase the luminosity of the cluster. This will tend to increase
the radiation pressure, and in turn the radius of the
cluster. However, as soon as $r$ begins to increase, the accretion
luminosity will drop so as to maintain the total luminosity near the
Eddington value. If the stellar luminosity exceeds the cluster
Eddington value, the stars may expel any remnant gas on a dynamical
time. 

However, the maximum stellar luminosity depends on the IMF, and is
generally well below the accretion luminosity, as we now show. The accretion
luminosity to mass ratio is
\be  
{L_{acc}\over \mcl}={4\pi Gc\over \kappa_{es}} {\kappa_{es}\over \kappa_d}.
\ee  
The first factor on the right hand side of this equation is the
Eddington light to mass ratio,
$(L/M)_{Edd}=6.6\times10^4\cm^2\s^{-3}$, while the opacity ratio
ranges from $0.1$ for solar metallicity dust (and $T\approx 100$K) to
$10$ for a low metallicity cluster. From Figure \ref{Fig:Jeans_IMF} or
\ref{Fig:L_over_M_Jeans_cl} the maximum stellar $L_*/M_*\approx 
2\times 10^4\cm^2\s^{-3}$, i.e., about $1/3$ of the Eddington $L/M$
ratio. Only for the most massive and metal rich clusters will the
stellar luminosity approach (or slightly exceed) the accretion
luminosity. However, the largest clusters actually have lower
temperatures, as we have just seen, so $\kappa_d$ is likely to be a
bit lower than the rough estimate $\kappa=3$ we used here. We conclude
that the stellar luminosity is lower than the accretion luminosity, at
least until most of the gas has turned into stars.

This assumes that no other mechanism can eject gas from the
cluster. One mechanism that might do so is a combination of
protostellar jets and stellar winds. Recall that the Jeans mass is of
order $5-10M_\odot$ in these clusters, so a substantial fraction of
stars will have massive winds. Both the protostellar winds and jets
will have mechanical luminosities less than a tenth of the stellar
Eddington luminosity. These winds and jets will shock, producing gas
at $T=10^9$K, but this gas will rapidly cool by conduction to
$T\approx 10^7$K. From that temperature it will cool radiatively in a
time of order $10^4$ years, less than a dynamical time, as we showed
above in \S \ref{sec: Cluster}. We tentatively
conclude that jets and winds will not expel the bulk of the cluster
gas.

Given that observed clusters are only a factor of a few times
larger than the minimum radius allowed by radiation pressure, it would
appear that the fraction of gas ejected may be of order $1/2$, but not
substantially larger. We leave this question for later work.

As with the optically thick accretion heated clusters discussed in \S
\ref{sec: thick}, the IMF of these clusters will be top heavy, with
Jeans masses ranging up to $10M_\odot$, as shown in
Fig. \ref{Fig:Jeans_Mass_cl}. When they are young, they will have
elevated light to mass ratios; when older, their mass to light ratios
will be higher than ordinary optically thin globular clusters of the
same age and metallicity; see Figure \ref{Fig:M_over_L vs M_cl} and
the next subsection.

The very high $m_J$ predicted by these models suggests that the lockup
fraction $\alpha(t)$ will be particularly small for these very massive
clusters, perhaps as small as $0.3$ or less
(Fig. \ref{Fig:lockup}). We have noted above that the gas expelled
from evolving stars is rapidly removed from present day globular clusters
\citep{2001ApJ...557L.105F}. The escape velocity from the more massive
clusters considered here is only slightly higher than in globulars, so
the same gas removal mechanism, whatever it is, should be
effective. The radius of a cluster of age $\tau$ will then depend on
the {\em present day} stellar mass as
\be   
\rcl(\tau,\mcl)=\alpha(\tau)^{-8/5}
\phi_{rad}
G^{1/5}
\left({\kappa\over 4\pi c}\right)^{2/5}
\mcl^{3/5}.
\ee   
This relation is shown as the dashed line in Figure \ref{Fig:size} for
$\alpha(t)\approx0.38$, appropriate for a $10$ Gyr old cluster with a Jeans mass
of $2M_\odot$.

If these massive clusters have larger Jeans masses than optically thin
globular clusters, they will experience a larger relative expansion in
$\rcl$ than the globular clusters. Evidence for this expansion may be
detectable in old massive clusters as a surrounding halo of stars
extending out to the tidal radius; both the larger relative expansion
and the larger number of stars in the massive clusters will render
this halo more visible than similar halos around globulars.

The small lockup fraction will also alter the present day values of
derived quantities such as surface density, relative to the initial
values give above. For example, the surface densities will be smaller by
$\alpha^3$, as illustrated by the dashed line in Figure \ref{Fig:Sigma}.

\subsection{The relation between luminosity and velocity dispersion in massive old clusters}
We noted above that the mass to light ratio $\Upsilon_V$ will be an
increasing function of cluster mass, even at fixed metallicity. This
will lead to a relation between luminosity and velocity dispersion
that differs from that expected from a simple application of the
virial theorem, $L\sim \sigma^5$.

We start with the observation that the luminosity of a cluster depends
on the IMF, or, in our case on the Jeans mass. Consider an old
cluster, in which the turnoff mass (the main sequence lifetime of a
star at the turnoff mass is equal to the age of the cluster) is
smaller than the Jeans mass. The mass to light ratio of such a cluster
is an increasing function of the Jeans mass; the cluster light is
coming from stars on the flat part of our adopted IMF, and the number
of such stars (at a fixed initial cluster mass) decreases with
increasing $\mj$. Another way to say this is that a larger $\mj$ leads
to more mass in massive stars, which contribute no light at late
times, leaving less mass in (and fewer numbers of) low mass stars to
illuminate the cluster's declining years. Hence
$\Upsilon_V=\mcli/L_V$, where the subscript $i$ denotes the initial
cluster mass, increases with increasing $\mj$. The mass to light ratio
for our IMF scales as $\Upsilon_V(\mj)\sim \mj^{0.73}$, as long as the
turnoff mass is below $\mj$.

The observationally accessible quantity is $\Upsilon_V(\mcl)$; we
measure the present day cluster mass, not the initial cluster
mass. For clusters with $T_{eff}>10$K (optically thick clusters with
sufficiently high accretion rates to heat the gas, but not to provide
radiation support) we can estimate the scaling $\Upsilon_V(\mcli)\sim
\mcli^{0.32}$. The exponent is $0.73$ times $7/16$, the latter arising
from the scaling relation between $\mj$ and the initial $\mcl$,
$\mj(\mcli)\sim \mcli^{7/16}$, found in \S\ref{sec: thick}. The
present day $\mcl$ is smaller than $\mcli$ by (at least) a
factor $\alpha$.  Given $\mcli$, we can calculate $\mj$ and
hence $\alpha$, whence we can find the present day $\mcl$. Doing so,
we find
\be  
\Upsilon_V(\mcl)\sim \mcl^{0.28}, 
\ee  
The situation for radiation supported clusters is similar, but the
Jeans mass varies more rapidly with the initial cluster mass,
$\mj\sim\mcli^{13/20}$. This occurs despite that fact that the
effective temperature of the cluster actually decreases with
increasing $\mcl$; the decreasing temperature is more than offset by
the rapidly increasing cluster radius (or decreasing density). The
result is $\Upsilon_V(\mcl)\sim\mcli^{0.47}$, and
\be 
\Upsilon_V(\mcl)\sim \mcl^{0.11}.
\ee 
%

%
%

Using the scaling of mass to light ratio $\Upsilon_V$ with cluster mass
$\mcl$, we can find the relation between the stellar luminosity and
the cluster velocity dispersion $\sigma\equiv v/\sqrt{\Gamma_{vir}}$,
where the constant $\Gamma_{vir}$ is defined by
$M_{vir}=\Gamma_{vir}\sigma^2 \rcl/G$. We start with
\be  
L_V = 5\times10^5 L_{V,\odot}
\left({2 \over \Upsilon_{V,6}}\right)
\left({\mcl\over 10^6M_\odot}\right)^{0.89},
\ee 
where $\Upsilon_{V,6}$ is the mass to light ratio for a cluster of
$10^6M_\odot$.

Combining this with equation (\ref{eqn: escape}) we find
\be   
L_V = 1.6\times10^6 L_{V,\odot}
\left({\sigma\over 20\kms}\right)^{4.45}
\left({\alpha(10{\rm Gyrs})\over 0.38}\right)^{4.45}
\left({2 \over \Upsilon_{V,6}}\right)
\ee   
for the present day luminosity of a radiation pressure supported
cluster formed $10$ Gyrs ago. For optically thick clusters that were
not radiation pressure supported, $L_V\sim\sigma^{3.6}$.

\section{Comparison with observed cluster properties}
\label{sec:compare}
We start by considering the light to mass ratio $L/M$ for young
clusters, then the mass to light ratio $M/L$ for old
clusters\footnote{We reverse the ratio with advancing age since this
  is the practice in the literature}. Either ratio is affected by a
number of parameters, including age, IMF, and metallicity. A cluster's
$L/M$ depends most strongly on age, moderately weakly on the IMF
(for the type of model considered in this paper), and very weakly on
metallicity. 

Figure \ref{Fig:M_over_L vs age} shows that $M/L$ varies by a factor
of several hundred over $10$ Gyrs for a given $\mcl$ and
metallicity. Figure \ref{Fig:M_over_L vs M_cl} shows that over the
observed range of cluster masses, variations in the Jeans mass should
produce a variation in the mass to light ratio (at 10 Gyrs) from
$\sim2 M_\odot/L_{V, \odot}$ at $\mcl=10^5M_\odot$ to $\sim6$ at
$\mcl=10^8M_\odot$, in increase of more than a factor of $3$. 

The metallicity of a cluster affects the $\Upsilon_V$ of a cluster with
a fixed IMF, since main sequence metal poor stars of a given mass tend
to be more luminous than metal rich stars of the same mass and age. 
However, the effect is not strong: Figure \ref{Fig:M_over_L vs M_cl}
shows that a factor of five change in metallicity changes $M/L_V$ (at
an age of 10Gyrs) by about $42\%$ around $\mcl=10^6M_\odot$, but by
only $20\%$ near $\mcl=10^8M_\odot$. Since the metallicity can be
obtained with moderate precision from spectroscopic studies of a
cluster, this direct source of variation in $M/L$ is reasonably easy
to account for.

The metallicity can also indirectly affect our estimate of $M/L$. For
example, \citet{jordan04} has pointed out that high metallicity
clusters can appear to have smaller half light radii than low
metallicity clusters of the same age, mass, and IMF; the mass of
turnoff stars (which dominate the light of the cluster) is smaller in
the metal poor cluster than that in the metal rich cluster.
The variation in radius is $\sim20\%$ for a factor of ten variation in
metallicity, and the change in the apparent mass is of the same order.

We conclude that clusters with $\mcl\approx 10^8M_\odot$ should have
$M/L_V\approx  6$ or higher, more than a factor of two higher than globular
clusters, and that metallicity variations cannot produce such a large
change in $M/L_V$. Armed with these results, we proceed to examine
various clusters for evidence of a top-heavy IMF.

\subsection{The light to mass ratio of young clusters}

In M82 the luminosities, ages, and projected half light radii of a
number of star clusters have been measured
\citep{SmithGallagher,McCrady03,Smith Westmoquett2006MNRAS.370..513S};
the first two papers also measured line of sight velocity dispersions
for a total of three clusters. The line of sight velocity dispersions
for 19 clusters have been measured by \citet{McCrady07}.

Both \citet{SmithGallagher} and \cite{McCrady03} present evidence that
cluster M82-F has a luminosity (in the F160W filter) to mass ratio
about a factor of $\sim2.5$ higher than expected for a cluster of its
estimated age ($\sim50$ Myrs). They suggest a lower mass cutoff to the
IMF of $2.5M_\odot$ and $\sim1M_\odot$, respectively. The projected
half light radius and mass of M82-F are $\rcl=89$ mas, about
$1.5\pc$ at $3.6$ Mpc, the distance to M82, and
$M_*=5.5\times10^5M_\odot$. Using these values, and a metallicity
$1.7$ times solar, we find that the Jeans mass in M82-F is
$\sim2.5M_\odot$.

The status of M82-F is the subject of some debate in the literature;
see \citet{2007MNRAS.379.1333B} for an extended discussion of this
object.  These authors find a high $\Upsilon$, as have previous
authors. They offer several possible explanations, including a
top-heavy IMF, but also mass segregation (not seen with their data)
and inaccurate estimates of the velocity dispersion resulting from
spatially variable extinction. They rule out the suggestion of
\citet{2006astro.ph..4464B} that M82-F is younger than $20$ Myrs.

Similarly, \cite{McCrady03} find that their cluster 11, with
$r_{hp}=1.2\pc$ and $M_*=3.5\times10^5M_\odot$, has $L/M$ about $2.5$
times larger than expected, again suggesting a top heavy IMF. Any mass
segregation in this cluster would have to be primordial, since the
cluster is so young, around 10 Myrs. For this cluster, our models
predict a Jeans mass of $2.3M_\odot$.

In contrast to these results for clusters M82-F and M82-11,
\cite{McCrady03} and \cite{McCrady07} find that their cluster 9
has $r_{hp}=2.6\pc$ and $M_*=2.3\times10^6M_\odot$. This cluster has
$L/M$ consistent with a normal IMF, while our models predict a large
Jeans mass and a high $L/M$ ratio.  

The other clusters in M82 are optically thin, so we do not expect them
to have elevated $L/M$ ratios.

Both \citet{SmithGallagher} and \citet{McCrady03} consider the effects
of mass segregation (heavier stars either forming preferentially or
settling into the cluster center). \citet{SmithGallagher} argue that
this is unlikely to explain the higher $L/M$, while \cite{McCrady03}
are more circumspect.

\citet{2005ApJ...620L..27B} find that the dynamically
evolving mass segregation will cause the half light radius to
decrease, weakening the argument for an enhanced light to mass
ratio in M82-F, but not in cluster 11 (due to its youth, dynamical
mass segregation is not important in this cluster).

Moving to young superstar clusters in galaxies other than M82, \citet{bastian06}
list eleven clusters less than 300 Myrs old; of these only NGC
4038:W99-15 is optically thick. Like M82:9, this cluster has a light
to mass ratio consistent with a normal IMF.  Other massive clusters
listed in \citet{bastian06}, such as NGC7252:W3, NGC7252:W30 and NGC
1316:G114, which are optically thick, are between 500Myr and 3Gyrs old
\citep{bastian06}, so they are also not expected to have elevated
$L/M$ ratios (see Fig. \ref{Fig:M_over_L vs age}).

To summarize, we are aware of only four young (less than 300Myr old)
clusters in the literature that were probably born optically thick to
far-infrared radiation, M82-F, M82-11, M82-9, and NGC 4038:W99-15. 
The first two show signs of having elevated light to mass ratios,
while the latter two do not.

\subsection{The mass to light ratio of massive star clusters}

Figure \ref{Fig:M_over_L vs M_cl} shows observed mass to light
ratios for several classes of star clusters, including globular
clusters from the Milky Way, M31, and Cen A, more massive UCDs
from the Virgo and Fornax clusters, and four nuclei of Virgo dwarf
ellipticals. Where available we have used masses from the literature
obtained by fitting models to individual clusters. Where such detailed
fits are not available, we have calculated the masses from the
observed velocity dispersions $\sigma$ and half light (or effective)
radii $\rcl$ using the expression
\be  
M=\Gamma_{vir}{\sigma^2\rcl\over G}.
\ee  
We use $\Gamma_{vir}=10$. We use only objects for which the quoted
error for $\sigma$ is less than half the value of $\sigma$.

Binning the clusters by mass ($\mcl<10^6M_\odot$,
$10^6<\mcl<10^7M_\odot$, and $10^7M_\odot<\mcl$), we find
$\Upsilon_V=1.9\pm0.8$, $2.8\pm1.2$, and $4.7\pm1.6$
respectively, all in solar units, consistent with the impression given
by the points representing individual objects. This rapid increase in
$\Upsilon$ with increasing cluster mass (or luminosity) has been noted
previously, e.g., by \citet{hasegan05}.

The mass to light ratio $\Upsilon$ of a cluster changes with age due
to both stellar evolution and to differential loss of low mass
compared to high mass stars. The latter is a result of two body
relaxation combined with tidal stripping, e.g.,
\citet{2003MNRAS.340..227B}.  As Fig. \ref{Fig:M_over_L vs age} shows,
the effect of stellar evolution is to increase $\Upsilon$ with
increasing cluster age. The solid curves in Fig. \ref{Fig:M_over_L vs
  M_cl} assume that the clusters are $10$ Gyrs old.  Some of the UCDs
may well be this young or younger; if they are younger, their
$\Upsilon$ values should be increased slightly to compare to the
globular clusters, since the latter almost certainly are $11-12$ Gyrs
old.

Metallicity also affects $\Upsilon$, but the majority of the objects
plotted in the figure, including the UCDs, have ${\rm [Fe/H]}<-1$. It is
worth stressing that for such low metallicities both the Kroupa and
Muench et al. IMFs predict $\Upsilon_V<3$ for these low metallicities.

\subsubsection{Dynamical effects}
Variations in $\Upsilon$ due to relaxation and tides are
most relevant for low mass objects.  Relaxation initially reduces
$\Upsilon$ as low mass stars are tidally stripped from the outskirts
of the cluster, then increases $\Upsilon$ just before the cluster is
disrupted at time $T_{dis}$. The disruption time found by the
numerical simulations is well approximated by
\citep{2003MNRAS.340..227B}
\be  \label{eq:diss}
T_{dis}=\beta\left[{N_*\over\ln(\gamma N_*)}\right]^x
{R_G\over\kpc}{220\kms\over v_c}(1-\varepsilon), 
\ee  
where the values of the constants are $\beta\approx1.9$,
$\gamma\approx0.02$, and $x\approx0.75$. The cluster is assumed to
orbit at a mean radius $R_G$ from the center of the galaxy on an orbit
with eccentricity $\varepsilon$. The galaxy has a circular velocity
$v_c$ (or a velocity dispersion which can be converted to an
equivalent circular velocity in the case of elliptical hosts). The
initial number of stars $N_*=\mcl/\langle m\rangle$, where $<m>$ is
the mean stellar mass for the chosen IMF.

The cluster mass at time $T$ is related to the initial cluster mass
$\mcli$ by
\be  \label{eq:relax} 
\mcl(T)=0.70\mcli(1-T/T_{dis}).
\ee  

From \citet{2003MNRAS.340..227B} (their Figure 14) we approximate
\be  \label{eq:ups} 
\Delta \Upsilon \approx \Gamma\left({T\over T_{dis}}\right),
\ee  
where $0.3\lesssim \Gamma\lesssim0.7$. From the observed cluster mass
(i.e., $\mcl(T)$) we solve equations (\ref{eq:diss}) and
(\ref{eq:relax}), then use equation (\ref{eq:ups}) to find
$\Delta\Upsilon$.  Where the eccentricity of the globular cluster
orbit is unknown (the majority of the cases) we assume
$\varepsilon=0.5$.

In producing Fig. \ref{Fig:M_over_L vs M_cl} we have increased the
mass to light ratio of all the clusters by the appropriate amounts,
using $\Gamma=0.7$; the only noticeable change (typically of order
$10-30\%$) is that suffered by the Milky Way globulars, since they are
the least massive and hence have the shortest $T_{dis}$. However, even
for Milky Way clusters the effect is not large, and neglecting this
correction does not alter the qualitative appearance of the figure.

There are recent high precision measurements of the present day mass
function (MF) in two globular clusters, M4$=$NGC 6121
\citep{2004AJ....127.2771R} and NGC 6397
\citep{2008AJ....135.2141R}. These were chosen for very long HST
observations partly based on the fact that they are the two nearest
globular clusters.

The metallicity of NGC 6397 is ${\rm [Fe/H]}=-2.03$
\citep{2003A&A...408..529G}, while the age of the cluster is $11.4$
Gyrs \citep{2008AJ....135.2141R}. Using $m_V=5.73$ and $r_h=2.33'$
\citep{Harris}, the error-weighted mean velocity dispersion 
$\sigma=3.5\pm0.2\kms$ \citep{pryor}, and distance modulus
$m-M=12.1\pm0.1$ or $D=2.6\pm0.1\kpc$ \citep{2008AJ....135.2141R}, the
dynamical mass to light ratio of NGC 6397 is $\Upsilon_V=1.6\pm0.2$ in
solar units. This is marginally consistent with a
Muench et al. stellar model with that age and metallicity,
($\Upsilon_V=1.96$) or for a Kroupa IMF ($\Upsilon_V=2.1$).

Using data from \citet{2004AJ....127.2771R} we find
$\Upsilon_V=1.2\pm0.1$ for M4, too low to be consistent with either
type of IMF, but consistent with a maximal amount of dynamical
evolution, i.e., $\Gamma=0.7$ and an age approaching the cluster
disruption age.

Applying the correction for preferential loss of low mass stars to NGC
6397 we find $\Upsilon_V=2.1$, while for M4 we find $\Upsilon_V=1.5$,
the latter now being marginally consistent with the Muench et
al. IMF. This suggests that M4 is more dynamically
evolved than our application of the \citep{2003MNRAS.340..227B} would
indicate.

The present day mass functions of both these clusters can be fitted by
a single power law $dN/dm\sim m^{-\alpha}$ with $\alpha=0.1$, compared
to the Salpeter value $2.35$ or the Muench et al. value $1.15$ (for
masses below the break mass). This is reminiscent of the findings of
\citet{2003MNRAS.340..227B}, who predict that the MF of evolved
clusters will be very flat. However, the simulations predict
$\alpha\approx0$ only when $90\%$ of the cluster lifetime has
passed. It seems somewhat unlikely that the first two clusters
examined (chosen for their proximity to us) should both be so near
their demise. The rather modest dynamical evolution needed to explain
the slightly low $\Upsilon_V$ for NGC 6397, coupled with the very low
$\alpha$, suggests some primordial mass segregation in that object.

\subsubsection{Non-baryonic dark matter?}
 \label{sec:DM}

The high $\Upsilon_V\sim6$ seen in some UCDs studied by
\citet{hasegan05} led them to suggest that their objects might contain
a mass in non-baryonic dark matter comparable to their stellar
mass. This suggestion is motivated by the following argument.  The
mean $\Upsilon_V\approx2.0\pm0.9$ for Milky Way globulars, using
dynamical masses corrected for the preferential loss of low mass stars
using eqns. (\ref{eq:diss}), (\ref{eq:relax}), and (\ref{eq:ups}) (as
noted above, uncorrected values of $\Upsilon_v$ are about $20\%$ lower
on average). In contrast, $\Upsilon_V\approx4.7\pm1.5$ for
$\mcl>10^7M_\odot$. As mentioned above, most of the more massive
objects have ${\rm [Fe/H]}<-1$, so neither Kroupa nor Muench et
al. IMFs can match the observations. Many of the UCDs appear to have
somewhat younger stellar populations than do the globulars, adding to
the difficulty.

Possible explanations for different values of $\Upsilon_V$ include
problems with the estimate of the mass to light ratios (e.g., poorly
measured velocity dispersions), dynamical effects such as the
preferential loss of low mass stars, changes in the IMF associated with
increasing cluster mass, or the presence of dynamically significant
amounts of non-baryonic dark matter in the more massive clusters. In
the latter case, the dark matter would have to have a mass $1.3$ times
larger than the baryonic matter inside $\rcl$ in cluster with
$\mcl>10^7M_\odot$, assuming an IMF that is the same as that in Milky
Way globular clusters. In other words, the massive clusters and UCDs
would have to be dark matter dominated inside $\sim10\pc$.

Given what little we know regarding dark matter, this is
unlikely, as we now show.

\begin{figure}
\plotone{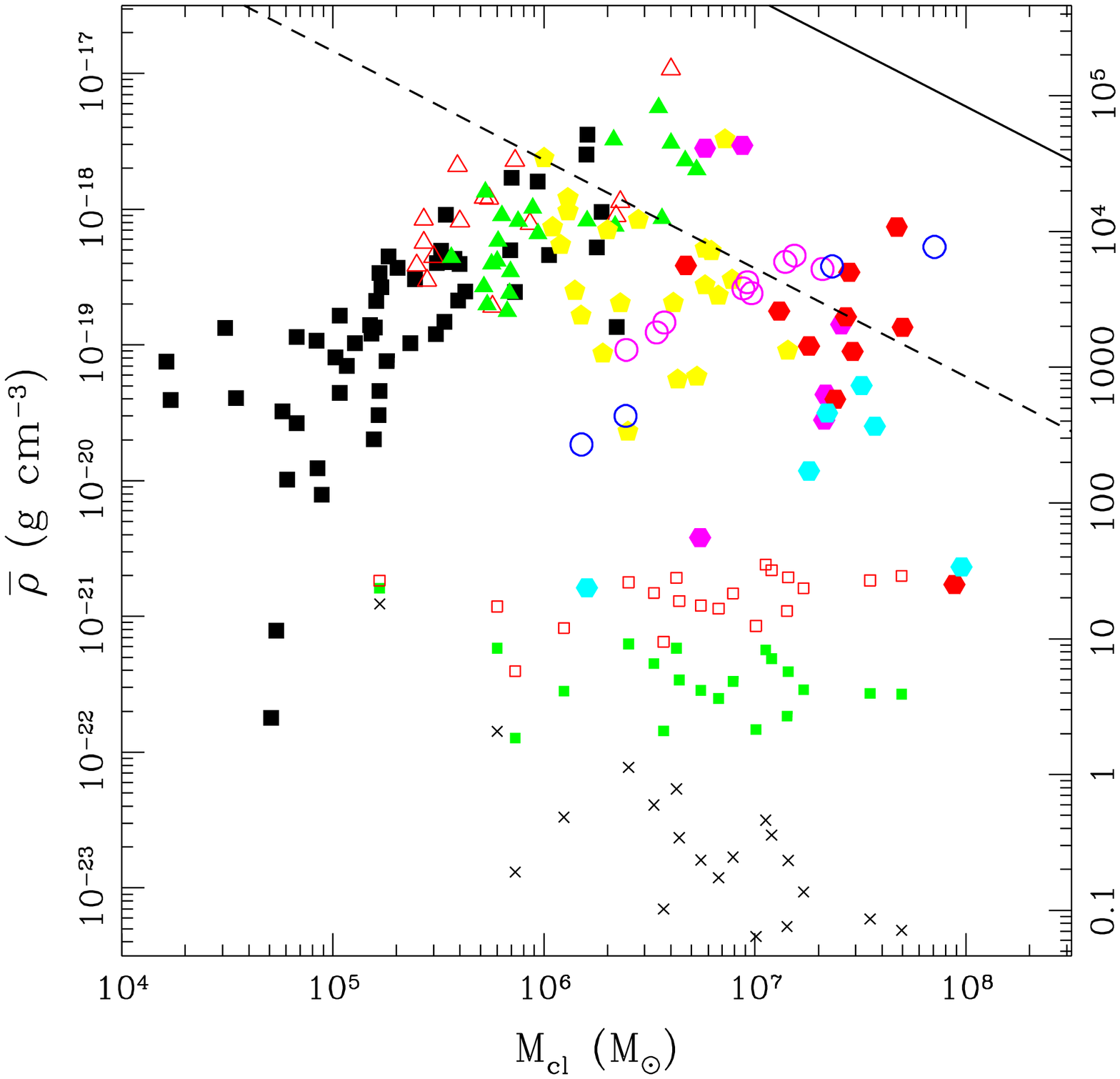}
\caption{
  Mean cluster density $\bar\rho = 3\mcl/(4\pi
  \rcl^3)$ as a function of cluster dynamical mass.  The symbols are
  as in Fig. 6, with the following additions: Open triangles are M82
  superstar clusters from \cite{McCrady07}. Open magenta circles are nuclear
  star clusers from \citet{walcher05}. Small crosses represent
  the mean density of Milky Way satellite dwarf galaxies evaluated at
  their core radii (typically substantially larger than $10\pc$). The
  small filled (green) and open (red) squares represent the satellite
  galaxy densities extrapolated to $10\pc$ using either an NFW profile
  or a Moore et al. (1999) profile. The solid line at the upper right
  represents the initial density of radiation pressure supported star
  clusters (assuming 100$\%$ star formation efficiency, while the
  dashed line shows the radius after stellar evolution has reduced the
  mass to $\alpha=0.4$ of its initial values. The right hand vertical
  axis is labeled by the density in solar masses per cubic
  parsec. Note that the dark matter dominated dwarf galaxies have mean
  densities at $10\pc$ a factor of (at least) $30$ less than the
  typical density of massive star clusters.
\label{Fig:density}}
\end{figure}

Figure \ref{Fig:density} shows the density of the same objects shown
in Fig. \ref{Fig:M_over_L vs M_cl}.

If non-baryonic dark matter is responsible for the elevated mass to
light ratios of clusters with $\mcl\sim10^7M_\odot$, it must have a
density $\gtrsim10^{-19}\g\cm^{-3}$, or $1000M_\odot\pc^{-3}$ on
scales of order $10\pc$. 

We have examples of dark matter dominated objects for which the
density is well measured, namely Milky Way dwarf spheroidal
galaxies. These objects have stellar velocity dispersions ranging from
$\sigma=4\kms$ to $27.5\kms$, the latter corresponding to the LMC. The
crosses in Fig. \ref{Fig:density} show the central densities
\be  
\rho_c = {166\sigma^2\over r_c^2}M_\odot\pc^{-3}
\ee  
of Milky Way dwarf spheroidal galaxies as tabulated by
\citet{madau}. In this expression $r_c=0.64\rcl$ is the core radius
of the stellar light (recall that $\rcl$ is the projected half light
radius).

The highest density Milky Way satellite galaxies currently known,
Willman 1 and Coma Berenices, have $\rho=10^{-21}\g\cm^{-3}$ and
$10^{-22}$ respectively, a factor of one hundred to one thousand below
the mean density of the compact $\mcl=10^7M_\odot$ clusters. More
massive dwarf galaxies have much lower mean (core) densities,
typically $\rho\sim 10^{-23}\g\cm^{-3}$; all these dark matter
densities are far to small to explain the high densities seen in
compact clusters.

The two highest density satellite galaxies (Willman 1 and Coma
Berenices) are also the most compact Milky Way galaxies, with
$r_c=13\pc$ and $41\pc$, respectively. This raises the possibility
that the larger satellites have higher densities at radii smaller than
$r_c$, as expected on theoretical grounds.  In fact, numerical
simulations find dark matter halos that follow a density profile given
by \cite{NFW}
\be  \label{eq:NFW}
\rho(r)={\rho_s\over (r/r_s) (1+r/r_s)^2},
\ee  
(the NFW profile) or the slightly steeper
\citet{1999MNRAS.310.1147M} profile
\be    \label{eq:Moore}
\rho(r)={\rho_s\over (r/r_s)^{1.5}(1+(r/r_s)^{1.5})}.
\ee    

If we assume $r_s>> r_c$ for the Milky Way dwarf satellites, we can
scale the density to $10\pc$; doing so, we find the small filled
squares in Fig. \ref{Fig:density} (NFW profile) and the small open
squares (the \citet{1999MNRAS.310.1147M} profile). The maximum mean
density at $10\pc$ is similar to that of Willman 1,
$\rho\approx10^{-21}\g\cm^{-3}$, far too small to explain the high
mass to light ratios of the massive star clusters and UCDs.

We show in appendix \ref{appendix:DM} that the relatively low inferred
mean dark matter density at $r=10\pc$ is consistent with the highest
resolution simulations, e.g., \citet{DKM}.

We conclude that compact massive star clusters (GCs and UCDs) are not
(non-baryonic) dark matter dominated. This does not mean that they
contain no non-baryonic dark matter: if they form in the center of
their own dark matter halo, the baryons will, when they collapse,
gravitationally compress the inner part of the dark matter halo
\citep{blumenthal}.  However, simple calculations show that the
fraction of dark matter inside $\rcl$ for objects as concentrated as
the compact clusters is typically less than $\sim30\%$.

\subsection{The mass-radius relation for massive
  clusters\label{sub:mass-radius}} 

\begin{figure}
\epsscale{1}
\plotone{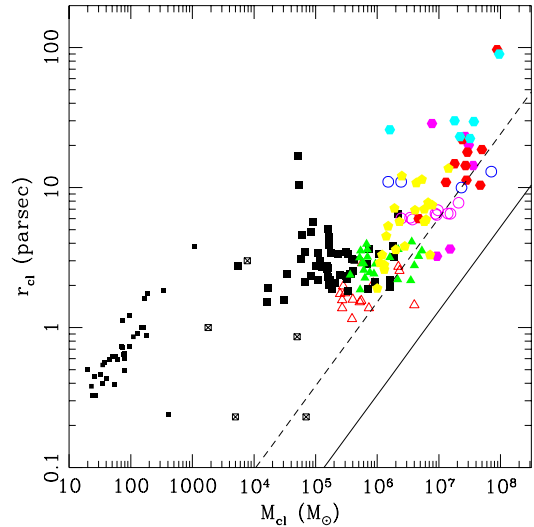}
\caption{
  Cluster radius in parsecs plotted against cluster mass (in
  solar masses). Symbols are as in Fig. \ref{Fig:density}, with two
  additions: Small filled squares are
  embedded (young) Milky Way clusters from \citet{L2}, while open squares
  with diagonals are massive young Milky Way clusters. The solid line
  shows the radiation radius computed from equation (\ref{eqn:
    radiation length mass}). The dashed line shows the cluster radius
  after mass loss (corresponding to a lockup fraction $\alpha=0.4$)
  assuming adiabatic expansion of the cluster.
\label{Fig:size} }
\end{figure}

We have already noted the striking observational result that both
young moderately massive ($10^4M_\odot<\mcl<10^6M_\odot$) clusters and
(old) globular clusters show no systematic variation of radius with
mass. Equation (\ref{eqn: radiation length mass}) predicts that the
mass of radiatively supported protoclusters should have radii that
increase with increasing mass. We therefore expect that evolved
massive clusters will inherit this mass-radius relation. Figure
\ref{Fig:size} shows the mass radius relation for low mass clusters in
the Milky Way, and high mass clusters and ultra compact dwarfs (UCDs)
from a variety of external galaxies. The prediction of equation (\ref{eqn:
  radiation length mass}) is shown as the solid line in the
Figure. This is the initial radius of a cluster, before it has
evolved, and should be compared with young massive clusters, such as
those in M82 studied by \citet{McCrady03} and \citet{McCrady07} (the
open triangles in the Figure). 

The dashed line is the predicted relation for old massive clusters;
the initial $\mcl$ has been reduced by the lockup fraction $\alpha$
(taken to be fixed at $0.4$, although it will vary with $\mcl$), while
$\rcl$ has been increased by the same factor. This line should be compared to
the old massive clusters and UCDs.

The fact that the low mass clusters ($\mcl\lesssim
5\times10^5M_\odot$) are well above the line indicates that something
other than radiation supported these clusters when they formed, i.e.,
that radiation support was not important in their evolution. In contrast,
the more massive clusters lie near the radius at which
radiation support becomes important, and their radii do show a trend
of increasing radius with increasing mass. This lends some credence to the
notion that radiation plays a role in the formation of the most
massive star clusters.

\begin{figure}
\plotone{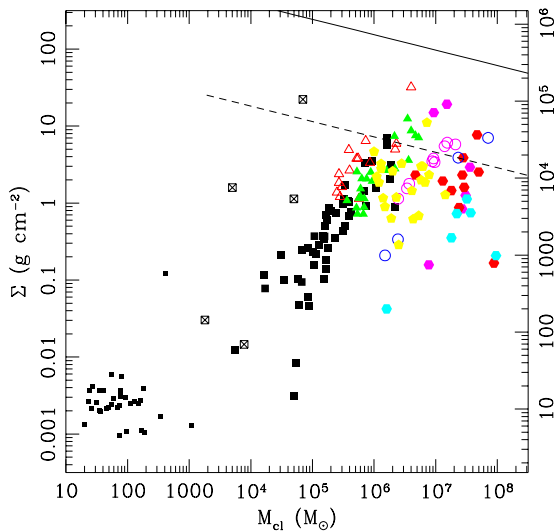}
\caption{
  Cluster surface density plotted against cluster mass (in solar
  masses). Data points are as in Fig. \ref{Fig:size}.  The solid line
  shows the maximum (initial) surface density allowed by the Eddington limit,
  equation (\ref{eqn:Sigma}). The dashed line shows the
  surface density after $\sim10$ Gyrs, with a lockup fraction
  $\alpha=0.4$. The right hand vertical axis is labeled in units of
  solar masses per square parsec.
\label{Fig:Sigma} }
\end{figure}

\section{DISCUSSION}
\label{sec:discussion}
The properties of star clusters with masses ranging from
$\sim100M_\odot$ to $10^8M_\odot$ vary continuously with cluster mass,
as seen in Figures \ref{Fig:M_over_L vs M_cl} through
\ref{Fig:Sigma}. There are sharp changes in the slope of cluster
radius at $\mcl\approx10^4M_\odot$ and at $\mcl\approx10^6M_\odot$,
and associated features in $\Sigma$, $\rho$ and cluster escape
velocity at these masses. 

We have argued that the change of slope in $\rcl(\mcl)$ at
$\mcl=10^6M_\odot$ arises from the emergence of radiation
pressure as a dynamically significant player in the formation of these
massive clusters. The feature at $\mcl=10^4M_\odot$ remains unexplained.

In contrast to the observation of two changes in the slope of $\rcl$
with $\mcl$, there is a single change in the slope of $\Upsilon_V$, at
$\mcl\approx10^6M_\odot$. We assert that this is due to the abrupt
increase in gas temperature with increasing cluster mass, associated
with the change from radiatively thin to radiatively thick
cooling. The sharp jump in $T(\mcl)$, the slower change of
$\rho(\mcl)$, and the stronger dependence of Jeans mass on $T$
($M_J\sim T^{3/2}/\rho^{1/2}$) results in a rapid increase in Jeans
mass with increasing cluster mass above $\mcl=10^6M_\odot$. We have
identified this as the underlying cause for the increase in
$\Upsilon_V$ with increasing $\mcl$ above $\sim 10^6M_\odot$.

The results regarding the IMF found here
(Figs. \ref{Fig:Jeans_Mass_cl}, \ref{Fig:M_over_L vs age}, and the
lines in Fig. \ref{Fig:M_over_L vs M_cl}) show that $\Upsilon_V$
should increase with increasing cluster mass even in the absence of
non-baryonic dark matter, up to values as large as $6$, and possibly
higher (the bulk of the extra mass is, in the case of a top heavy IMF,
in the form of white dwarfs). We have also argued that non-baryonic
dark matter in Galactic satellite galaxies on scales of
$\sim10\pc$ has densities smaller by a factor of 30 than the densities
observed in massive star clusters and UCDs. It appears unlikely that
the high mass to light ratios of the compact clusters are due to
non-baryonic dark matter.

\citet{FK02} model UCDs as merged globular clusters. Their clusters,
at $\mcl=2\times10^7M_\odot$ and with radii ranging from $39$ to
$72\pc$ or larger, are larger than all but one of the objects
plotted in Figure \ref{Fig:size}. 

\citet{2001ApJ...552L.105B} and \citet{2003MNRAS.344..399B} argue that
UCDs are the remnants of nucleated dwarf galaxies; the nuclei lose
their envelopes due to tidal stripping in the gravitational field of
their host cluster halos. This begs the question of how the nuclei
formed.  To address this question, \citet{2004ApJ...610L..13B} explore
the same mechanism as \citet{FK02}, merging of globular clusters. They
find scaling relations that imply a mass radius relation
\be  
\rcl\sim \mcl^{0.38},
\ee  
which would imply, for example, that the surface density of massive
clusters would increase with increasing mass. This would appear to be
ruled out for clusters with $\mcl\gtrsim 3\times10^6M_\odot$. 

Globular clusters have mass to light ratios $\Upsilon_V\approx 2$,
consistent with the notion that they contain no non-baryonic dark
matter. Models for massive clusters that merge globulars will not
naturally trap large masses of dark matter, so the merger products
will also lack dark matter, and have $\Upsilon_V\approx 2$, smaller
than the values observed for many of the UCDs shown in Figure
\ref{Fig:M_over_L vs M_cl}.

\citet{evstigneeva07} find from spectroscopy that the
ages, metallicities, and abundances of Virgo UCDs are similar to those
of Virgo globular clusters. However, the mass to light ratios and the
mass-radius relations differ. This is seen most clearly in the lower
panel of their Figure 6, which is very roughly a plot of $\Sigma$
against mass (as in Figure \ref{Fig:Sigma} here). They conclude that
the internal properties of Virgo UCDs are consistent with them ``being
the high-mass/high-luminosity extreme of known GC populations'',
apparently emphasizing the continuity of their distribution in the
$\kappa_2-\kappa_1$ (roughly the $\Sigma-\mcl$) fundamental plane
rather than the clear break in the slope that they find. 

In this work the break in scaling properties is attributed to the
emergence of radiation pressure as a dynamically significant element
in the formation of the cluster. This result strengthens the case for
treating globular clusters and UCDs on a unified footing.

\citet{MK} note the high values of $\Upsilon_V$ found for UCDs, and
suggest that it is due to a non-standard IMF. Their explanation is that
the IMF is bottom-heavy, i.e., that there are more low mass stars per
unit stellar mass than in the standard (Milky Way) IMF. It may appear
paradoxical that high $\Upsilon$ values arise from both top heavy and
from bottom heavy IMFs, but in both cases one appeals to an excess of
under-luminous stars. In a top heavy IMF, at late times, the extra
dead weight is found in stellar remnants, while in bottom heavy IMFs,
the dead weight is found in low mass stars (well below the main
sequence turnoff). \cite{MK} suggest observations of CO at
$2.3\mu{\rm m}$ as a way to test for the presence of large numbers of
low mass stars.

\citet{2007MNRAS.374L..29K} suggest that in warm star-bursting
circumnuclear gas, as in the Galactic center, the IMF will be top
heavy compared to the IMF rest of the Milky Way. They present
numerical simulations showing that this is the case. They attribute
the difference to the larger Jeans mass in the warm gas, as argued
here. In this paper, we attribute the higher temperature to accretion
of the cluster gas, as opposed to radiation from stars surrounding the
proto-cluster; however, the crucial point is that higher temperatures in the
ISM will lead to a top heavy IMF.

\section{CONCLUSIONS}
\label{sec:conclusions}
We have shown that the Jeans mass in a cluster is roughly independent
of the cluster mass as long as the cluster is optically thin to the
FIR, and assuming that the mass radius relation seen in embedded
clusters reflects the primordial mass radius relation. Clusters
forming today in the Milky War are optically thin, so this finding is
consistent with the observed constancy of the IMF in our galaxy.

Many Milky Way globular clusters were also optically thin, or had low
enough velocities that their accretion luminosities would not have raised
the gas temperature significantly. The Jeans mass in such clusters
would then depend only on their radii at the time of
formation. 

However, we have pointed out that many clusters in other galaxies, and
in the Milky Way in the past, are or were optically thick to the FIR;
such clusters often (though not always) have high enough accretion
luminosities that the gas temperature would have been higher than
$10$K. We went on to argue that the Jeans mass is larger than a solar
mass in such clusters. Using the assumption that the break in the IMF
is associated with the Jeans mass, we concluded that massive clusters
should have high $L/M$ ratios when young, and high $M/L$ ratios when
more than a few Gyrs old.

The prediction of high $L/M$ is consistent with observations of
apparently enhanced light to mass ratios in the superstar clusters
M82-F and M82-11, although the optically thick clusters M82-9 and NGC
4038:W99-15 appear to have normal $M/L$ ratios. Other superstar
clusters in M82 and other nearby galaxies are optically thin, and so
they should have normal IMFs. Clearly, more data on
superstar clusters would be helpful.

A top heavy IMF, which we showed should occur in the most massive and
compact globular clusters, will tend to produce an excess of LMXBs and
pulsars in metal rich globular clusters in both the Milky Way and in
nearby galaxies, and at the same time, a higher $M/L$ ratio. 

Compelling support for the notion of a high mass to light ratio in
massive clusters is supplied by Figure \ref{Fig:M_over_L vs M_cl},
which compares the predicted and observed mass to light ratio for
objects with masses ranging up to $10^8M_\odot$. As the solid lines in
the Figure indicate, this result is consistent with the prediction of
a top-heavy IMF resulting from the elevated temperatures associated
with accretion in an optically thick environment. 

We also argued that in the most massive clusters, with $\mcl\gtrsim
3\times10^6M_\odot$, the substantial luminosity associated with the
contraction of the cluster must have been dynamically important. The
predicted mass-radius relation, $\rcl\sim \mcl^{3/5}$, leads to a
number of relations between the properties of massive star clusters,
involving the cluster velocity dispersion, the cluster luminosity, and
the surface density; for example, we found $L_V\sim
\sigma^4$. These relations appear to be consistent with the
observed properties of massive star clusters.

The combination of strong evidence for a cluster-mass dependent mass
to light ratio and for a mass-radius relation $\rcl\sim\mcl^{3/5}$
provides solid support for the idea that a contraction powered
radiation field has left its mark on the most massive star clusters we
see around us.

It is fitting to close by emphasizing again the striking fact that
globular clusters are observed to have radii of $2-3$ parsecs, with
only weak dependence on cluster mass over a range $10^4M_\odot$ to
$3\times10^6M_\odot$. The origin of this (lack of a) mass-radius
relation remains a major puzzle.

\acknowledgments  
The author is grateful to Natasha Ivanova for helpful
discussions.  This research has made use of the SIMBAD database,
operated at CDS, Strasbourg, France, and of NASA's Astrophysics Data
System.  The author is supported in part by the Canada Research Chair
program and by NSERC of Canada.

\appendix
\section{LOG-NORMAL PROBABILITY DENSITY FUNCTIONS}
 \label{appendix:lognormal}
Both analytic theory and numerical simulations of turbulent flows find
that the density follows a log-normal probability density
function. While the interpretation of observations is difficult, there
are indications that gas in star forming regions in the Milky Way does
as well. In this appendix we show that this fact implies that the
characteristic density $\rho_m$ of the gas out of which stars in star
clusters form is no more than a factor of $\sim10$ larger than the
mean density of the protocluster.

Following \citet{osg01}, let $y=\log(\rho/\bar\rho)$, where the
logarithm is base 10 and $\bar\rho$ is the mean density of the
cluster. Then the probability density function is
\be  
f_M(y)={1\over\sqrt{2\pi \sigma^2}}
\exp\left[-(y-\mu)^2/(2\sigma^2)\right].
\ee  
The quantity $f_M(y)dy$ is the fraction of the cluster mass with
density contrast $y$ in the interval $(y,y+dy)$. Demanding
conservation of mass (in the form of the continuity equation) leads to
a relation between the mean $\mu$ and the dispersion $\sigma$ of $y$:
\be  
\mu = {1\over2}\ln(10)\sigma^2.
\ee  

Letting
\be  
t=(y+\mu)/\sqrt{2}\sigma,
\ee  
the fraction of the cluster mass having $\rho>\rho_m$ is
\be  \label{eqn: lognormal} 
f_g(>\rho_m)={1\over2}{\rm erfc}(t_m).
\ee  

\begin{figure}
\plotone{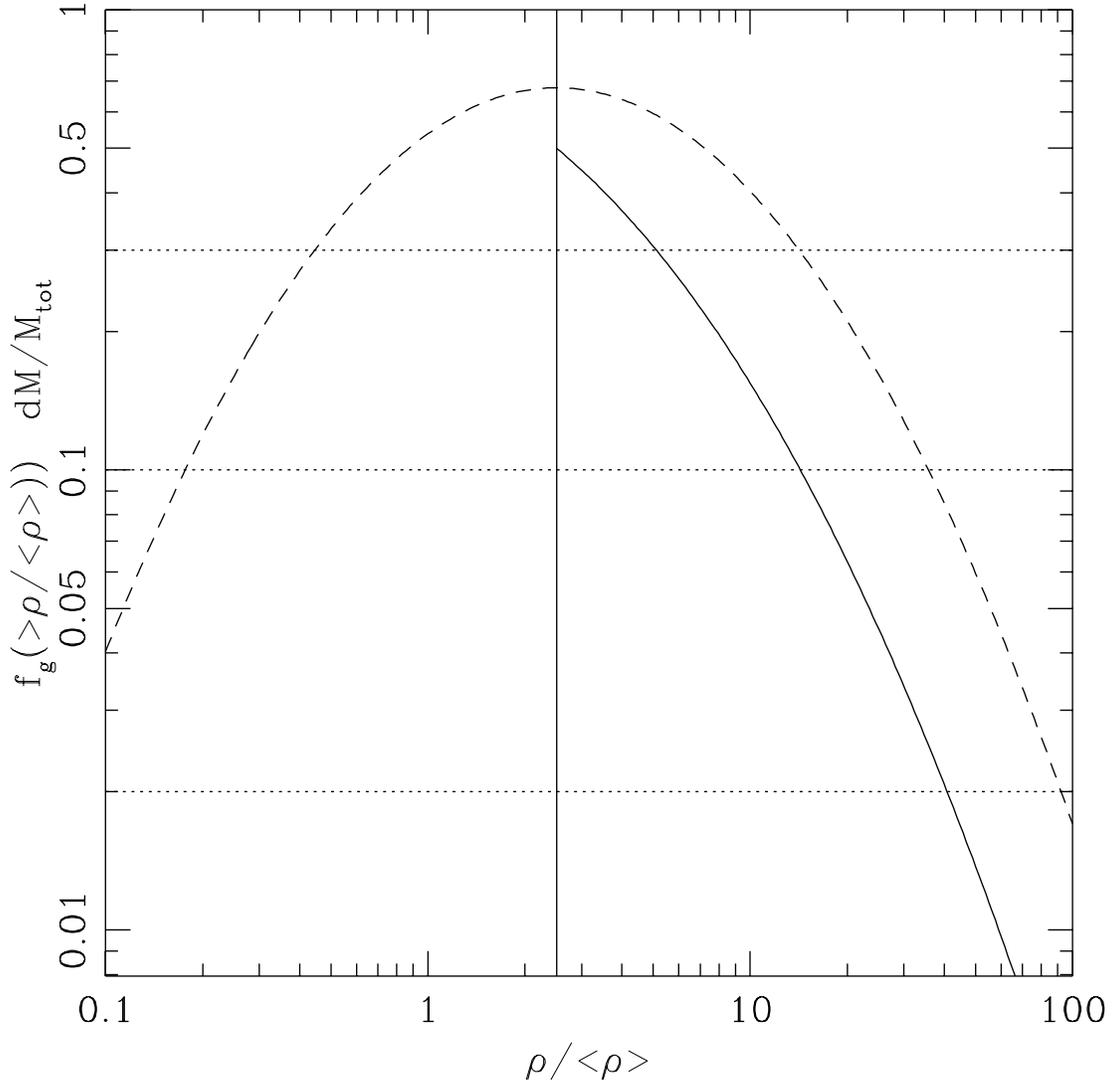}
\caption{
The solid line shows the gas fraction $f_g(>\rho/\bar\rho)$ from
eqn. \ref{eqn: lognormal}.  Also shown is the probability density
function $f_M(y)=dM/M_{tot}$ (dashed line) for $\mu\equiv \langle \log
\rho/\bar\rho \rangle_M =0.4$ (corresponding to ${\cal M}_F\approx3$). The solid
vertical line is at the mass-weighted mean density $\mu$.  The
horizontal dotted lines are drawn assuming that $f_g=SFE$, i.e., that
only the densest gas in the cluster eventually ends up in stars, for
$SFE=0.3$, $0.1$, and $0.1/5$. The value last assumes that the stars
form over 5 dynamical times, so that $f_M$ is sampled five times.
\label{Fig: lognormal} }

\end{figure}

Figure \ref{Fig: lognormal} plots both $f_M(\rho/\bar\rho)$ and
$f_g(>\rho_m)$. We have taken $\mu=0.4$; in the simulations of
\citet{osg01} this corresponds to a fast magnetosonic Mach number
$M_F\approx3$. 

If star formation occurs on a crossing time, the probability
density function is sampled once before the protocluster gas is
dissipated. During this time a fraction between $0.1$ and $0.3$ of the
gas must form stars. These fractions are indicated by the upper two
horizontal dotted lines in the Figure. The corresponding over
densities are $5$ and $14$ times the mean density of the clump. Under
the extreme assumption that only the densest gas in the protocluster
forms stars, the Jeans mass of the relevant gas is factor between two and four
smaller than the mean Jeans mass of the clump.

If star formation takes more than a single crossing time, the
probability density is sampled more than once, so a larger fraction of
the gas may be turbulently compressed to high density. The bottom
dotted line corresponds to star formation in a cluster in which
$SFE=0.1$ and for which star formation persisted for five dynamical
times. We see that the characteristic density for this case is a
factor $40$ larger than the mean density, resulting in a Jeans mass
$6.3$ times smaller than the Jeans mass calculated using the mean
density.

These estimates for the Jeans mass are likely underestimates; many
self-gravitating bodies of gas with lower density but larger sizes
will be produced by the turbulence, as argued by, e.g., \citet{PN}. In
fact the argument made here is similar to that made by those authors;
both rely on the steep decline in the PDF with increasing
density. Here we are emphasizing the observational constraints
provided by the SFE and estimates of the duration of star formation
rather than the dynamical constraint that stars form out of self
gravitating material.

\section{LUMINOSITY FUNCTIONS AND INITIAL MASS FUNCTIONS}
\label{appendix:imf}
The number of stars in a cluster having a given mass is described by
the initial mass function (IMF), $\phi(m)dm$, the number of stars with
masses between $m$ and $m+dm$, where $m$ is measured in solar masses
$M_\odot=2\times10^{33}\g$. The IMF is normalized so that
\be \label{eq: imf}
\int_{m_L}^{m_U}m\phi(m)dm=1,
\ee 
where $m_L$ and $m_U$ are lower and upper mass limits. We sometimes employ
the Salpeter mass function, which for $m_L=0.1$, $m_U=100$, is
$\phi_{Sal}=0.17m^{-\alpha}$ with $\alpha=2.35$.  We also use the observed IMF for the
Orion Nebula, given in Muench et al. (2002); for simplicity we set
$\phi(m)=0$ for masses below $0.025M_\odot$.

%
\be \label{eq: muench}
\phi_{M,m_1}(m)\equiv{dN\over dm} = N_0\left\{ \begin{array}{ll}
m^{-2.21}     & m_U>m>m_1\\
m_1^{1.15-2.21} m^{-1.15}     & m_1>m>m_2\\
m_1^{1.15-2.21} m_2^{0.27 - 1.15} m^{-0.27}     & m_2>m>m_L
\end{array}
\right.
\ee  
%
\citet{2002ApJ...573..366M} 
found $m_1=0.6M_\odot$ and $m_2=0.12M_\odot$. The high mass end of the
IMF is described by a power law with index $\alpha=2.21$, similar to
that of the Salpeter IMF. Note that we use $dN/dm$
rather than $dN/d\log m$, so the exponents that appear here are equal
to those quoted in \citep{2002ApJ...573..366M} minus one, i.e., their
exponent $-1.21$ becomes our exponent $-2.21$, while their $+0.73$
becomes $-0.27$ here. 

The Muench et al. IMF is similar to that of 
\citet{2001MNRAS.322..231K}, which has a high mass slope of
$\alpha=2.3$ and a low mass slope $\alpha=1.3$, slightly steeper than
that of Muench et al.

%
%

\subsection{The light to mass ratio\label{appendix:mass_to_light}}
We use the Padova stellar evolution tracks (\cite{PadovaI};
\cite{PadovaII}; Girardi 2006, http://pleiadi.pd.astro.it) to find the
luminosity of stars of mass $m$ at an age of $\approx 2.5$ Myrs. For
stars with $m<9$ we use the isochrone for $4$ Myrs; these stars
contribute very little luminosity, so the error involved in doing so
is minor. For more massive stars we searched the evolutionary tracks
to find the age nearest $2.5$ Myrs,
since the luminosity of these stars varies rapidly with age. The
result is $L(m)$ in the form of a table.

To find the light to mass ratio for stars at an age of 2.5Myrs, we
integrate 
\be \label{eq: L_over_M}
\int_{m_L}^{m_U}L(m)\phi(m)dm.
\ee 
For a Salpeter IMF with $m_L=0.1$ and $m_U=120M_\odot$ (the estimated
initial mass of $\eta$ Carina) the light to mass ratio is $2010$ in
cgs units. Using the Muench et al. (2002) IMF with the $m_L$ and
$m_U$, $m_2=0.1$ and $m_1=0.6$ the ratio is $4470$. Part of this
difference is due to the difference in slope at the high mass end of
the two IMFs; using a slope of 2.21, the light to mass ratio of a
simple powerlaw is 1790. The rest of the difference is due to what is
effectively a low mass cutoff at $m=0.6$ for the Muench IMF. This
factor of two difference in the light to mass ratio for the different
IMFs will lead to a factor of two difference in the predicted
efficiency of star formation.

In the main text we argue that $m_1$ is set by the Jeans mass, and
that the latter varies with cluster mass. Figure \ref{Fig:Jeans_IMF}
shows the light to mass ratio for a Muench et al. IMF as a function of 
$m_1$, keeping $m_L$, $m_U$, and $m_2$ fixed at their original values.

\section{RADIATION PRESSURE SUPPORTED CLUSTER SIZES}
 \label{appendix: size}

While there is strong observational support for a characteristic
globular cluster size, we are not aware of a good physical explanation
for it, nor do we give one here. Instead, in this appendix we
investigate the possibility that for massive clusters,the cluster
length scale is determined by the interplay between the self-gravity
of a clump of gas, the turbulent pressure $\rho v_T^2$, and in very
dense and massive systems, radiation pressure. The turbulent pressure
is
\be   
P_{turb}=\rho v_T^2.
\ee   
The turbulent velocity is well above the sound speed in the clusters
under consideration here, so we expect that it will decay on the
dynamical time of the cluster. The turbulent pressure therefor cannot
halt the collapse of the protocluster, but should act to slow it. We
assume that as the collapse proceeds, the gravitational binding energy
released is converted into new turbulent motions, which then shock and
dissipate their energy as heat. This heat is converted into thermal
radiation, which then diffuses out of the cluster.

To model this, we assume that the turbulent pressure is a fixed
fraction of the dynamical pressure, or, in terms of accelerations
\be   
a_{turb}=(1-\gamma) |a_{grav}|= (1-\gamma){GM(r)\rho\over r^2}.
\ee   
We usually take $\gamma\approx  0.2$; choosing different values leads to
different cluster radii and Jeans masses, when the protocluster is
radiatively supported.

The thermal radiation will diffuse out of the protocluster, but before
it does it provides a pressure opposed to that of gravity:
\be   
P_{rad}={a\over3}T^4,
\ee   
where $a=4\sigma_B/c$ and $\sigma_B$ is the Stefan-Boltzman constant.
In the optically thick limit one can model this as a diffusion
process, so that 
\be   
P_{rad}= {\kappa(T,\rho,Z) \rho L(r)\over 4\pi r^2 c},
\ee   
where $\kappa(T,\rho,Z)$ is the Rossland mean opacity for the
appropriate temperature, density, and metallicity. In our crude
models we assume that the luminosity is roughly constant, independent
of radius, as would be the case for a flow that is near free fall. 

It is easy to show that for these massive clusters the gas pressure is
negligible compared to the radiation pressure; one way to see this is
note that the turbulent motions are highly supersonic.

The momentum equation for an infalling shell is
\be   
{dv\over dt}= -\gamma {GM(r)\over r^2} + {\kappa L\over 4\pi r^2 c}.
\ee   
The luminosity is given by
\be  
L={GMM\over r^2} v(t).
\ee  
Using this we can write
\be  
{dv\over dt}=-{GM(r)\over r^2} 
\left[ 
\gamma - {\kappa\over c} 
{M(r)\over 4\pi r^2 }v
\right].
\ee  
The first term on the right hand side of this equation is the
effective (turbulence reduced) acceleration due to gravity, while the
second term is the acceleration due to radiation pressure. The
radiation term is proportional to the rate of collapse.

In our numerical work we start the integration at a large radius,
where the second term is negligible. We stop the integration when the
radiation pressure slows the collapse by $20-30\%$ compared to the
case with no radiation pressure. More exactly, if
$v_g\equiv\sqrt{G\mcl/r}$, we define $\tau_g\equiv r/v_g$ and $\tau_{dyn} =
r/v$. We stop the calculation when $\tau/\tau_g > 1.2$ or $1.3$. Using
the approximation that $v\approx \sqrt{2\gamma GM(r)/r}$, this occurs
at a radius
\be   
r_{rad}=
\phi_{rad}
G^{1/5}
\left({\kappa\over 4\pi c}\right)^{2/5}
M^{3/5},
\ee   
where $\phi_{rad}$ is a
dimensionless constant; for our simple model, $\phi_{rad}\approx 3$.

The rational for stopping the integration is that, when the dynamical
time $\tau_{dyn}$ exceeds the gravitational crossing time $\tau_t$,
the cluster is partially supported by radiation pressure. The rate at
which gravitational binding energy is converted into turbulent motion
is reduced, so the level of turbulent pressure support will drop; the
turbulent dissipation time will remain comparable to $\tau_{grav}$, so
the turbulence will decay more rapidly than the cluster shrinks.  We
assume that as the turbulence decays, star formation will proceed at a
rapid pace, and the cluster radius will be frozen in at or about the
radius at which radiation pressure support becomes important.

\section{NUMERICAL ESTIMATES OF DARK MATTER
  DENSITIES} \label{appendix:DM} 

In section \S \ref{sec:DM} we argued by reference to the properties of
known dark matter-dominated objects (Milky Way satellite dwarf
galaxies) that dark matter densities on the scale of $10\pc$ were
$\rho_{DM}\sim10^{-21}\g\cm^{-3}$, much less than the densities of
massive compact star clusters ($\rho\sim10^{-19}\g\cm^{-3}$: see
Fig.~\ref{Fig:density}). In this appendix we use the results of
numerical calculations to estimate the maximum density on small
scales \citep{NFW,2001MNRAS.321..559B,DKM,madau}.

It is traditional to use the concentration
$c_{vir}\equiv r_{vir}/r_s$, where 
\be  
r_{vir}\equiv
\left(
{3 M_{vir}\over 4\pi \Delta_{vir} \rho_{crit}}
\right)^{1/3}
\ee  
is the virial radius of a dark matter halo, and $r_s$ is the
characteristic radius of the density distribution, introduced in
equations (\ref{eq:NFW}) and (\ref{eq:Moore}). In this expression
$\rho_{crit}\equiv 3H_0^2/(8\pi G)\approx 9\times10^{-30}\g\cm^{-3}$
is the critical density of the universe and $\Delta_{vir}\approx200$,
depending on the cosmology, e.g., $\Delta_{vir}=180$ for an
Einstein-de Sitter cosmology. 

We consider an NFW profile; results for \citet{2001MNRAS.321..559B}
are similar. The mean density of an NFW halo at radius $r$ is
\be  
\bar\rho(r)= 3 \Delta_{vir}\rho_{crit}
\left( {r_{vir}\over r} \right)^3
{A(r/r_s)\over A(c_{vir})},
\ee  
where
\be  
A(x) = \ln(1+x) - {x\over 1+x}.
\ee  

In the limit that $r<<r_s$ this is
\be  
\bar\rho(r)\approx 
1500 {c_{vir}^2\over A(c_{vir})}
\left({M_{vir}\over 10^8M_\odot}\right)^{1/3}
\left({10\pc\over r}\right)
\Delta_{vir}\rho_{crit},
\ee  
where we have scaled to the minimum virial mass for a cluster baryon
mass of $10^7M_\odot$. The value of $c_{vir}\approx10$ for a Milky Way
size halo ($M_{vir}\approx10^{12}M_\odot$), leading to an estimate of
$\bar\rho(10\pc)\approx4\times10^{-21}\g\cm^{-3}$. From the results
reported in \citet{DKM} and \citet{madau}, the concentration
$c_{vir}\lesssim50$ for $M_{vir}\approx10^8M_\odot$ for subhalos near
the center of their parent halo (note that both Fornax and Virgo UCDs
are close to the centers of their parent halos). For these low mass
compact halos, we find
$\bar\rho(10\pc)\approx3\times10^{-21}\g\cm^{-3}$, slightly higher
than the observed densities of the Milky Way satellite galaxies (when
extrapolated to $r=10\pc$), but not nearly high enough to be
dynamically important in the massive compact clusters and UCDs.

We note that the numerical calculations performed to date are not of
high enough resolution to measure the density on scales of
$10\pc$. The highest resolution study published to date, that of
\citet{DKM}, has a force resolution of $\sim90\pc$, sufficient to
resolve clusters $\sim300\pc$ in radius. Since we have no theoretical
understanding of the density profiles found in the simulations, the
extrapolation from $300\pc$ to $10\pc$ we use to estimate the dark
matter densities is on shaky ground. For this reason we prefer the
direct comparison with objects such as Willman I, where the dark
matter density is measured directly on the relevant scale.


\end{document}